\documentclass{article}

\usepackage[english]{babel}
\usepackage[utf8x]{inputenc}
\usepackage[T1]{fontenc}
\usepackage[title]{appendix}
\usepackage{framed}
\usepackage{color}
\usepackage{algorithm}
\usepackage{algpseudocode}
\usepackage{float}
\usepackage{enumerate}
\usepackage{enumitem}
\usepackage[a4paper,top=3cm,bottom=2cm,left=3cm,right=3cm,marginparwidth=1.75cm]{geometry}
\usepackage{amsmath}
\usepackage{amssymb}
\usepackage{amsthm}
\usepackage{proof}
\usepackage{graphicx}
\usepackage{subcaption}
\usepackage[colorinlistoftodos]{todonotes}
\usepackage[colorlinks=true, allcolors=blue]{hyperref}
\usepackage{wrapfig,lipsum}
\usepackage{authblk}
\usepackage{tikz}
\usetikzlibrary{matrix, shapes, arrows,topaths,calc}
\usepackage{multirow}
\usepackage{hhline}
\usepackage{setspace}
\usepackage{palatino}
\usepackage{mathdots}
\usepackage{multirow}
\usepackage{siunitx}
\usepackage{booktabs,colortbl, array}
\usepackage[normalem]{ulem}
\PassOptionsToPackage{svgnames}{xcolor}
\usepackage{listings}
\usepackage{tabularx}
\pgfdeclarelayer{background}
\pgfdeclarelayer{foreground}
\pgfsetlayers{background,main,foreground}

\newcolumntype{C}[1]{>{\centering\arraybackslash}p{#1}}

\newcommand\ket[1]{\ensuremath{|#1\rangle}}
\newcommand\bra[1]{\ensuremath{\langle#1|}}

\newcommand\Tr{\mathop{\rm Tr}\nolimits}

\newtheorem{definition}{Definition}

\begin{document}

\title{Alibaba Cloud Quantum Development Platform: Surface Code Simulations with Crosstalk}

\author[1]{Cupjin Huang\thanks{cupjin.huang@alibaba-inc.com}}
\author[2]{Xiaotong Ni}
\author[3]{Fang Zhang}
\author[4]{Michael Newman}
\author[1]{Dawei Ding}
\author[1]{Xun Gao}
\author[2]{Tenghui Wang}
\author[5]{Hui-Hai Zhao}
\author[2]{Feng Wu}
\author[2]{Gengyan Zhang}
\author[2]{Chunqing Deng}
\author[2]{Hsiang-Sheng Ku}
\author[1]{Jianxin Chen} 
\author[1]{Yaoyun Shi}

\affil[1]{Alibaba Quantum Laboratory, Alibaba Group USA, Bellevue, WA 98004, USA}
\affil[2]{Alibaba Quantum Laboratory, Alibaba Group, Hangzhou, Zhejiang 311121, P.R.China}
\affil[3]{Department of Electrical Engineering and Computer Science, University of Michigan, Ann Arbor, MI 48109, USA}
\affil[4]{Departments of Physics and Electrical and Computer Engineering, Duke University, Durham, NC 27708, USA}
\affil[5]{Alibaba Quantum Laboratory, Alibaba Group, Beijing 100102, P.R.China}
\setcounter{Maxaffil}{0}
\renewcommand\Affilfont{\itshape\small}

\date{\today}
\maketitle

\begin{abstract}
\noindent
We report, in a sequence of notes, our work on the Alibaba Cloud Quantum Development Platform (AC-QDP). AC-QDP provides a set of tools for aiding the development of both quantum computing algorithms and quantum processors, and is powered by a large-scale classical simulator deployed on Alibaba Cloud.  
In this note, we simulate a distance-3 logical qubit encoded in the 17-qubit surface code using experimental noise parameters for transmon qubits in a planar circuit QED architecture. Our simulation features crosstalk induced by ZZ-interactions. We show that at current noise levels,  crosstalk contributes significantly to the dephasing of the logical qubit. This results in a total phase-flip probability of $\sim 0.6\%$, about $60\%$ higher than expected without considering crosstalk. Our results quantify the significant damage caused by simple, correctable crosstalk, emphasizing the need to mitigate even its second-order effects in order to realize beneficial error-correction.
\end{abstract}

\medskip

\section{Introduction}

It is generally accepted that scalable quantum computation requires quantum error-correction.  Although fully fault-tolerant quantum computing has been firmly established in theory \cite{aliferis2005quantum, knill1996threshold, gottesman1998theory, dennis2002topological}, it has yet to be realized in practice.  In the interim, there have been increasing demonstrations of elements of quantum error-correction, including classical error-correction in a quantum system \cite{kelly2015state}, quantum error-detection \cite{corcoles2015demonstration, linke2017fault}, and improvements in qubit lifetime via bosonic encodings \cite{ofek2016extending}.  Thus, an important milestone for each computing platform is the demonstration of full fault-tolerance.  While expected to be reached in the near future, this task is challenging due to the relatively large number of physical qubits and high precision of operations required.

Typically, resource requirements for quantum error-correction are estimated using efficient simulation within the stabilizer formalism \cite{gottesman1998heisenberg}. However, these simulations are far-removed from the different physical sources of noise and their realistic effects.  As quantum simulators become increasingly available, there have been several full-scale simulations of small quantum error-correction experiments \cite{obrien2017density, tomita2014low, darmawan2017tensor, bermudez2017assessing}.

Complementary to advances in physical realizations, these classical simulations can be instrumental for yielding physically-motivated, accurate estimates of device requirements to realize beneficial error-correction.  In addition, they can shorten the device development cycle. The process of designing, fabricating, and calibrating a superconducting quantum processor can take weeks. Classical simulators can provide a rapid and cost-effective tool to guide the choices of the many device parameters.

Classical simulation has its own set of challenges. Given that the simulation complexity is necessarily exponential, the primary constraint is the scale, which calls for highly efficient algorithms and implementations. 
In addition, even when performing general noise simulation, there are inaccuracies that fall into the following two major categories.
\begin{itemize}
    \item Modeling errors. For superconducting circuits and many other systems, we usually truncate each mode to a finite-dimensional Hilbert space and then use the resulting Hamiltonians along with dissipation terms to describe the dynamics. Such models may not be accurate enough to describe the actual dynamics, while some other noise processes may be unaccounted for altogether. 
     \item Discretization errors. Once we have modeled the system, we need to compute the time evolution. If it is modeled by Hamiltonians and dissipation terms, then the time evolution is computed by solving a differential equation. In general, this is harder to simulate than a quantum circuit in the gate model. A standard method to solve these differential equations is to discretize the time steps and use a Trotter approximation. However, this can potentially yield a very large overhead, as the depth of the circuit can increase significantly. The necessary tradeoffs made for computational feasibility may cause substantial errors.
\end{itemize}
In this work, we study the effects of certain noise processes and possible ways to reduce them.
In particular, we focus on crosstalk noise in a specific superconducting circuit implementation of surface-17, a small instance of the surface code family~\cite{bravyi1998quantum, bombin2007optimal} (see~\autoref{fig_surface}).
While the effects of different types of noise in surface-17 do not necessarily represent the behavior of larger surface codes, it is likely that noise reduction techniques learned from simulating small systems can be applied to larger ones.

In order to support our simulation, we developed a quantum error correction module for Alibaba Cloud Quantum Development Platform (AC-QDP). The computational engine of AC-QDP is a refactored version of Tai-Zhang \cite{CZH+18}, which not only enables the simulation of $\ge50$ qubit quantum circuits on the cloud, but also provides verification and benchmarking tools for noisy intermediate-size ($50-100$ qubits) quantum devices. 
The design and the benchmark results of Tai-Zhang are reported in~\cite{CZH+18} and~\cite{ZHN+19}. AC-QDP can also be used to assist the design and testing of quantum algorithms~\cite{HSZ+19, S19}.

\section{Experimental Setup}

\subsection{Surface-17}

\begin{figure}[hbt!]
    \centering
    \begin{tikzpicture}[auto, scale=1.5]
        \tikzstyle{data} = [regular polygon,regular polygon sides=4,draw=none,fill=blue]
        \tikzstyle{anx} = [circle,draw=none,fill=red]
        \tikzstyle{anz} = [circle,draw=none,fill=green]

        \draw[fill=green!20, draw=none] (1,0) circle (1);
        \draw[fill=green!20, draw=none] (3,4) circle (1);
        \draw[fill=red!20, draw=none] (4,1) circle (1);
        \draw[fill=red!20, draw=none] (0,3) circle (1);
        \draw[fill=green!20, draw=none]  (2,4) -- (0,4) -- (0,2) -- (2,2) -- cycle;
        \draw[fill=red!20, draw=none]  (2,2) -- (4,2) -- (4,4) -- (2,4) -- cycle;
        \draw[fill=green!20, draw=none]  (2,2) -- (4,2) -- (4,0) -- (2,0) -- cycle;
        \draw[fill=red!20, draw=none]  (2,2) -- (0,2) -- (0,0) -- (2,0)-- cycle;

        \node[data, label=below:{(0, 0)}] (00d) at (0,0) {};
        \node[data, label=below:{(0, 2)}] (02d) at (0,2) {};
        \node[data, label=below:{(0, 4)}] (04d) at (0,4) {};
        \node[data, label=below:{(2, 0)}] (20d) at (2,0) {};
        \node[data, label=below:{(2, 2)}] (22d) at (2,2) {};
        \node[data, label=below:{(2, 4)}] (24d) at (2,4) {};
        \node[data, label=below:{(4, 0)}] (40d) at (4,0) {};
        \node[data, label=below:{(4, 2)}] (42d) at (4,2) {};
        \node[data, label=below:{(4, 4)}] (44d) at (4,4) {};

        \node[anx, label=below:{(1, 1)}] (11x) at (1,1) {};
        \node[anx, label=below:{(5, 1)}] (51x) at (5,1) {};
        \node[anx, label=below:{(-1, 3)}] (-13x) at (-1,3) {};
        \node[anx, label=below:{(3, 3)}] (33x) at (3,3) {};

        \node[anz, label=below:{(1, 3)}] (13z) at (1,3) {};
        \node[anz, label=below:{(1, -1)}] (1-1z) at (1,-1) {};
        \node[anz, label=below:{(3, 1)}] (31z) at (3,1) {};
        \node[anz, label=below:{(3, 5)}] (35z) at (3,5) {};

        \draw (00d) to (11x);
        \draw (02d) to (11x);
        \draw (20d) to (11x);
        \draw (22d) to (11x);
        \draw (22d) to (33x);
        \draw (24d) to (33x);
        \draw (42d) to (33x);
        \draw (44d) to (33x);
        \draw (02d) to (13z);
        \draw (22d) to (13z);
        \draw (04d) to (13z);
        \draw (24d) to (13z);
        \draw (20d) to (31z);
        \draw (40d) to (31z);
        \draw (22d) to (31z);
        \draw (42d) to (31z);
        \draw (20d) to (1-1z);
        \draw (00d) to (1-1z);
        \draw (44d) to (35z);
        \draw (24d) to (35z);
        \draw (40d) to (51x);
        \draw (42d) to (51x);
        \draw (02d) to (-13x);
        \draw (04d) to (-13x);

        \end{tikzpicture}
    \caption{Illustration of surface-17. The blue squares, red circles, and green circles represent the data, X-, and Z-ancilla qubits, respectively. The shaded red and green regions represent the X- and Z-type stabilizers, respectively. Two-qubit interactions only exist between connected qubit pairs.}
    \label{fig_surface}
\end{figure}
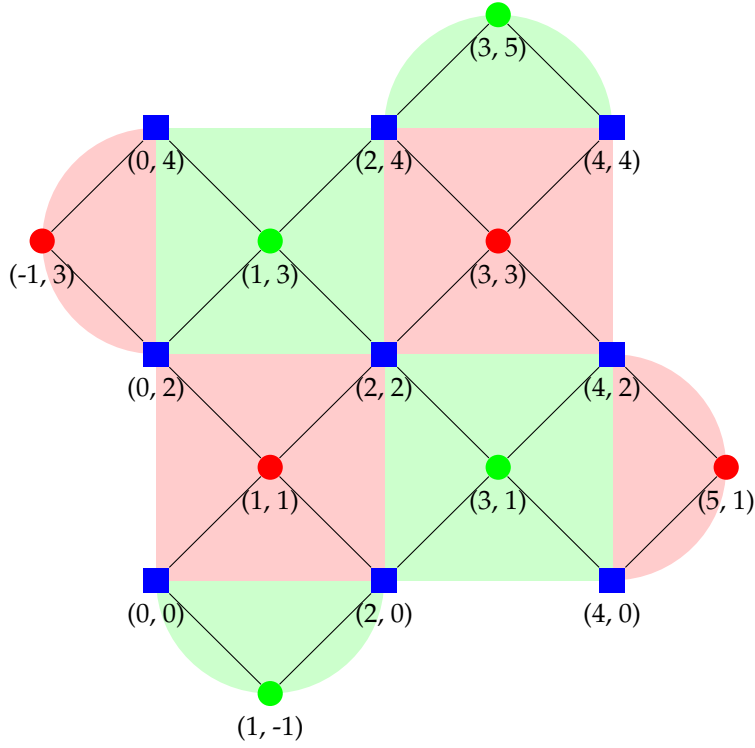

A surface code is a stabilizer code defined on a square grid. Here we consider the surface code defined on a $3\times 3$ grid, which consists of $9$ physical qubits. To realize the code in practice, ancilla qubits must be introduced in order to extract the classical syndromes used for error correction. The $9$ physical qubits, together with $4$ ancilla qubits for $X$ stabilizer measurements and $4$ ancilla qubits for $Z$ stabilizer measurements form the 17 qubit surface code, illustrated in \autoref{fig_surface}.

\subsection{Logical Memory Experiment}
In this section, we report the simulation results for a surface-$17$ simulation at the level of the logical quantum memory. Each experiment consists of $k$ rounds of noisy syndrome extraction, and one round of noiseless syndrome extraction together with a trivial decoder for this round.
The goal of the trivial decoder is to perform simple corrections so that the final state lies in the code space.
In particular, we choose a weight-1 pure error (see Chapter 6.3 of~\cite{lidar2013quantum}) for each stabilizer we measure.
If the measurement outcome of a stabilizer in the final round is $-1$, we noiselessly apply the corresponding pure error to correct it.
The exact assignments of pure errors are not important because we will perform the actual decoding, which is insensitive to these assignments, afterwards.
As a result, we have a channel which leaves the code space invariant, and to each syndrome we must assign a logical Pauli correction. The full logical channel is then obtained by choosing the optimal logical Pauli correction conditioned on each syndrome, and summing up over all syndromes.

For simplicity, we focus on the channel restricted to the code space. We use the initial state 
$$\ket{\Psi} = \ket{0}_{\text{logical}}\ket{0}_{\text{extra}} + \ket{1}_{\text{logical}}\ket{1}_{\text{extra}} , $$
which is a maximally entangled state between the logical qubit and an additional virtual qubit.
The extra qubit is only used to extract the logical channel via the Choi–Jamiołkowski isomorphism. Note that the resulting simulation is unphysical in the sense that it does not represent a real experiment, but nonetheless gives information quantifying how damaging different errors can be.
Also note that the optimal decoder is used instead of more efficient decoders such as those based on minimum-weight perfect matching~\cite{dennis2002topological, fowler2012surface} or union-find algorithms~\cite{delfosse2017linear, delfosse2017almost, li20192d, huang2019fault, das2020scalable}. The reason for our choice is two-fold. First, optimal decoders, despite being ultimately unscalable, are within reach for experiments on surface-17. Second, using the optimal decoder in the simulation allows us to focus on the quantum hardware to determine the effects of each hardware parameter on the performance of quantum error correction, so that the results can best guide the demonstration of a fault-tolerance pseudothreshold.

\subsection{Pauli Transition Matrix}
Instead of focusing on a specific quantity characterizing the error rate, we report the whole logical channel in the form of a Pauli transition matrix (PTM). The PTM of a single-qubit channel $\mathcal{C}$ is a $4\times 4$ real matrix defined as 
$$P(\mathcal{C})_{ij}=\Tr[\sigma_i\mathcal{C}(\sigma_j)],$$
where $\sigma_0, \sigma_1,\sigma_2,\sigma_3=I,X,Y,Z$, respectively. Note that the first row of any PTM corresponding to a completely positive and trace preserving (CPTP) map is always $(1,0,0,0)$ since $\Tr[\sigma_0\mathcal{C}(\sigma_j)]=\Tr[\mathcal{C}(\sigma_j)]=\Tr[\sigma_j]=\begin{cases}
    1 & \text{if } j = 0;\\
    0 & \text{if } j = 1,2,3.\\
    \end{cases}$ The other 12 entries in the matrix represent the degrees of freedom, where roughly, the first column indicates non-unital shifts, the diagonal terms indicate dephasing errors and the off-diagonal terms indicate coherent errors (See \autoref{fig_ptm}). Moreover, the phase-flip error rate $p_{\text{phase}}$ and bit-flip error rate $p_{\text{bit}}$ can be directly deduced from the corresponding PTM entries.

\begin{figure}[hbt!]
    \centering
    \begin{tikzpicture}[every node/.style={anchor=base, text depth=0.5ex,text height=2em,text width=8em}]
        \matrix [inner sep=2mm]
        {
        \node[fill={rgb:black,1;white,2},inner sep=2.8mm] {\ \ \ \ \ \ \ \ \ \ 1}; & \node[fill={rgb:black,1;white,2},inner sep=2.8mm] {\ \ \ \ \ \ \ \ \ \ 0}; & \node[fill={rgb:black,1;white,2},inner sep=2.8mm] {\ \ \ \ \ \ \ \ \ \ 0}; & \node[fill={rgb:black,1;white,2},inner sep=2.8mm] {\ \ \ \ \ \ \ \ \ \ 0}; \\
        \node[fill=blue!30!white,inner sep=2.8mm] {X shift}; & \node[fill=white!90!green,inner sep=2.8mm] {$1-2p_{\text{phase}}$}; & \node[fill=yellow!30!white,inner sep=2.8mm,inner sep=2.8mm] {Z rotation}; & \node[fill=yellow!30!white,inner sep=2.8mm] {Y rotation}; \\
        \node[fill=blue!30!white,inner sep=2.8mm] {Y shift}; & \node[fill=yellow!30!white,inner sep=2.8mm] {Z rotation}; & \node[fill=white!90!green,inner sep=2.8mm,inner sep=2.8mm] {1-$2\Pr[\text{Y error}]$}; & \node[fill=yellow!30!white,inner sep=2.8mm] {X rotation}; \\
        \node[fill=blue!30!white,inner sep=2.8mm] {Z shift}; & \node[fill=yellow!30!white,inner sep=2.8mm] {Y rotation}; & \node[fill=yellow!30!white,inner sep=2.8mm,inner sep=2.8mm] {X rotation}; & \node[fill=white!90!green,inner sep=2.8mm] {$1-2p_{\text{bit}}$}; \\
        };
        \end{tikzpicture}
    \caption{Illustration of a single-qubit PTM}
    \label{fig_ptm}
\end{figure}

\section{Experimental Error Model}\label{sec:model}

To achieve a faithful simulation,  one needs to use the Lindblad equation to compute the time evolution of the density matrix, including the cavities needed for measurements.
For tensor-network simulation, we will have to consider a discretized version of time evolution, which inevitably introduces inaccuracies into the simulation.
Considering there may be additional modeling errors, how accurate our simulation would be in predicting experimental results is yet to be examined. Even if the errors are significant, a
qualitative understanding of the effects of crosstalk may still be achieved.

\begin{figure}[ht!]
    \centering
    \begin{subfigure}{\textwidth}
        \centering
        \includegraphics[width=\textwidth]{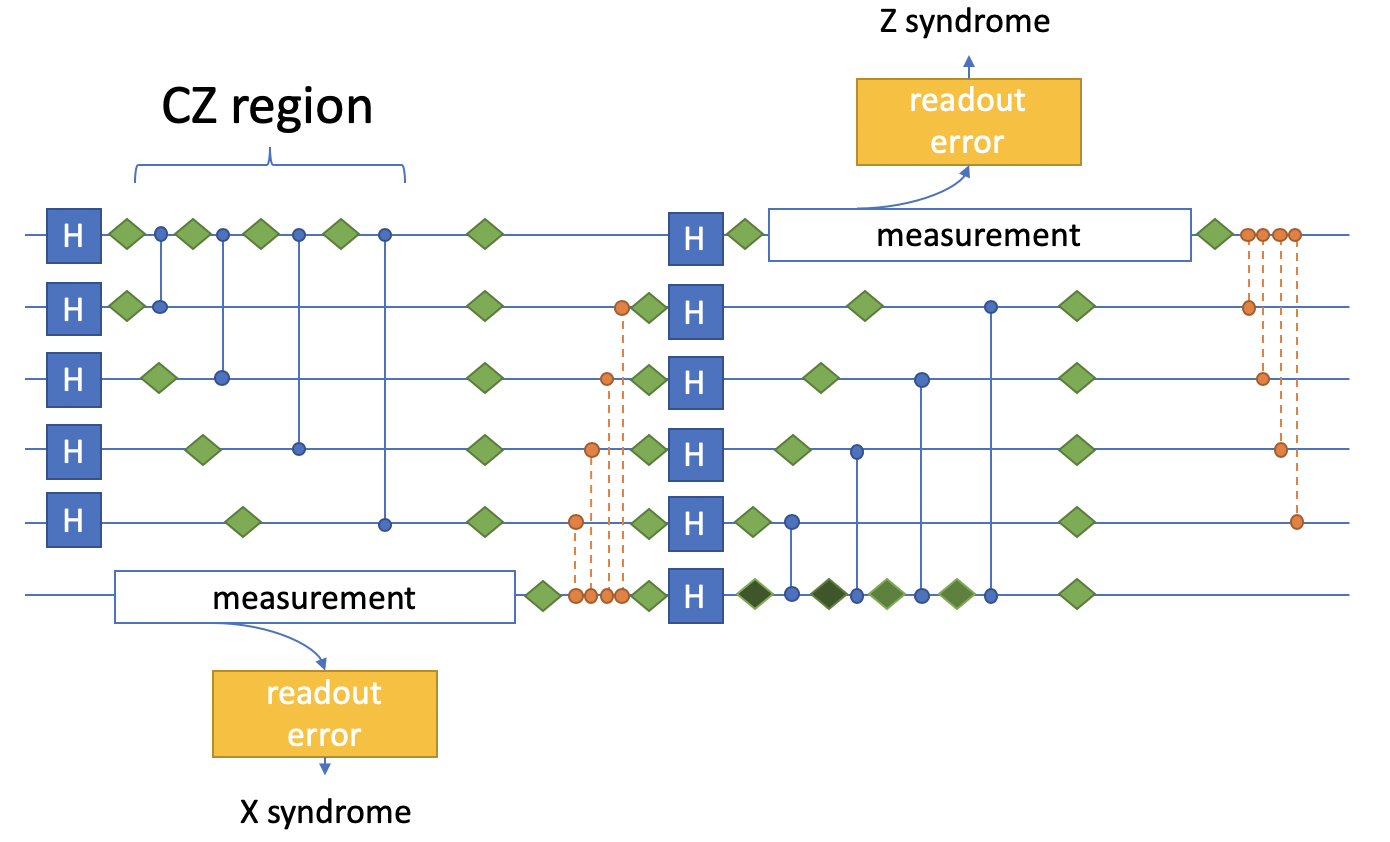}
        \caption{The gate-based noise model. The noise can be roughly classified into four parts: the idling errors (green), the gate-specific errors (blue), the readout errors (yellow) and crosstalk (orange). The dark green diamonds represent amplitude-phase damping channels with an additional phase damping from photon decay. This diagram does not reflect the actual model in the sense that no four data qubits are adjacent to the same X- and Z-type ancilla qubits simultaneously, but we present it this way for the sake of simplicity.}
    \end{subfigure}
    \newline
    \vspace*{0.3cm}    
    \newline
    \begin{subfigure}{0.3\textwidth}
        \centering
        \includegraphics[width=\textwidth]{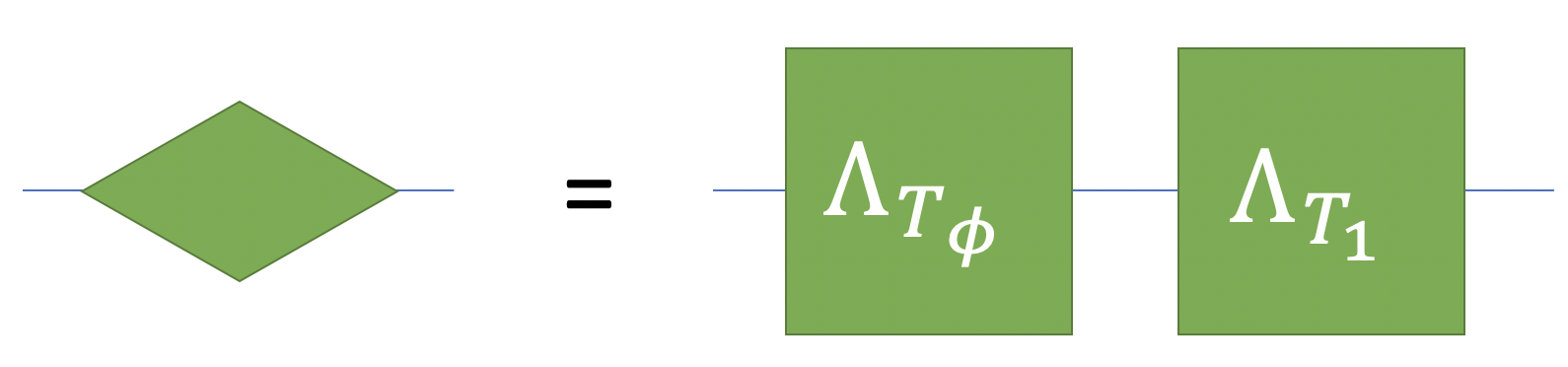}
        \caption{Idling errors}
    \end{subfigure}
    \hfill
    \begin{subfigure}{0.3\textwidth}
        \centering
        \includegraphics[width=\textwidth]{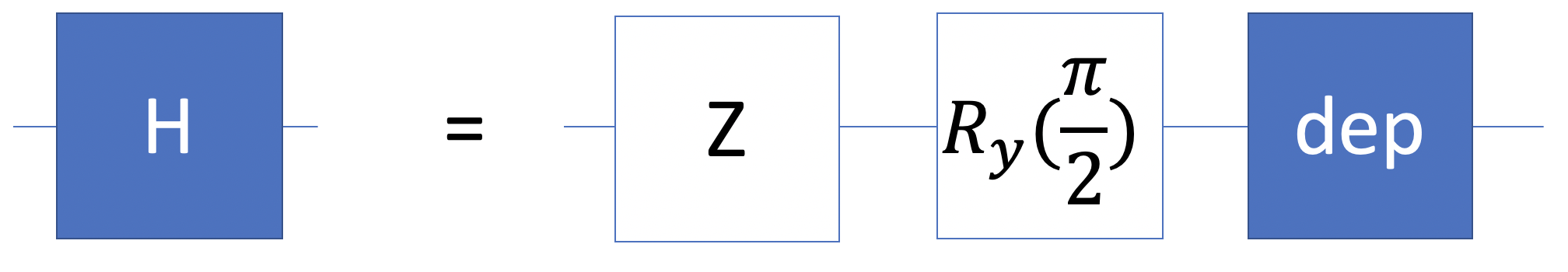}
        \caption{Noisy Hadamard gates}
    \end{subfigure}
    \hfill
    \begin{subfigure}{0.3\textwidth}
        \centering
        \includegraphics[width=\textwidth]{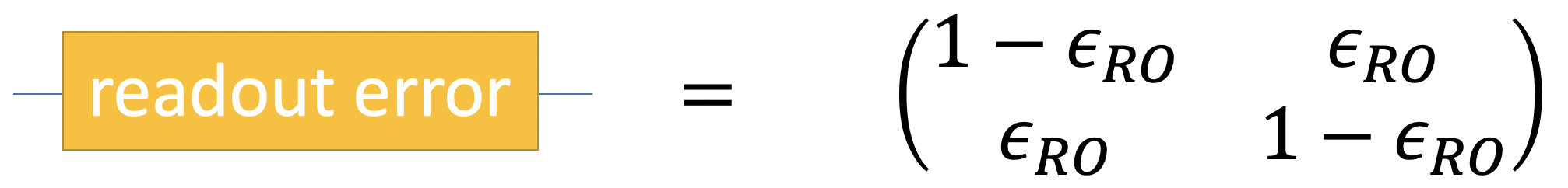}
        \caption{Readout error channels}
    \end{subfigure}
    \caption{An illustration of the gate-based noise model}
    \label{fig_full}
\end{figure}

For the sake of completeness, we introduce our gate-based noise model which is based on the model presented in \cite{obrien2017density}. See \autoref{fig_full} for an illustration of the model. Whereas \cite{obrien2017density} uses Pauli transfer matrices to represent error channels, we will use Choi matrices, as they are closer to the tensor representations used in our program.

\subsection{Idle error}\label{subsec:idle}
The error on an idling qubit is described by the standard amplitude-phase damping model, which as the name suggests, consists of two components:
\begin{itemize}
    \item Amplitude damping, which causes a qubit $A$ to relax from the excited state $\ket{1}_A$ to the ground state $\ket{0}_A$. When idling for some time $t$, the excited state has a probability $p_1$ to emit a photon and decay to the ground state. We can assume that the environment electromagnetic field is initialized in the vacuum state $\ket{0}_E$, and the emission of a photon is a unitary process that changes the environment to $\ket{1}_E$.
    
    This evolution can be thus described as follows:
    \begin{itemize}
        \item Adding an ancillary qubit $\ket{0}_E$.
        \item Apply a unitary  such that 
        $$\ket{00}_{AE}]\mapsto\ket{00}_{AE}\quad\textrm{and}\quad \ket{10}_{AE}\mapsto \sqrt{1-p_1}\ket{10}_{AE} + \sqrt{p_1}\ket{01}_{AE}.$$
        \item Discard the ancilla.
    \end{itemize}
    \item Phase damping, which causes the qubit to dephase into the computational basis $\{\ket{0}_A, \ket{1}_A\}$. This occurs due to interactions that are weak compared with the system energy but strong compared with the environment energy. These are not strong enough to cause the qubit to flip from the excited state to the ground state or vice versa, but may cause the environment state to change depending on the qubit state.
    
    This evolution can be described as follows:
    \begin{itemize}
        \item Adding an ancillary qubit $\ket{0}_E$.
        \item Apply a unitary  such that
         $$\ket{00}_{AE}\mapsto \ket{00}_{AE} \quad\textrm{and} \quad\ket{10}_{AE}\mapsto\sqrt{1-p_\phi}\ket{10}_{AE} + \sqrt{p_\phi}\ket{11}_{AE}.$$
        \item Discard the ancilla.
    \end{itemize}
\end{itemize}

In the Choi matrix representation:
$$C_{\Lambda_{T_1}}=\left[
\begin{array}{cc|cc}
    1 & 0 & 0 & \sqrt{1-p_1} \\
    0 & 0 & 0 & 0 \\
    \hline
    0 & 0 & p_1 & 0 \\
    \sqrt{1-p_1} & 0 & 0 & 1 - p_1
\end{array}
\right]\qquad\textrm{and}\quad
C_{\Lambda_{T_\phi}}=\left[
\begin{array}{cc|cc}
    1 & 0 & 0 & \sqrt{1-p_\phi} \\
    0 & 0 & 0 & 0 \\
    \hline
    0 & 0 & 0 & 0 \\
    \sqrt{1-p_\phi} & 0 & 0 & 1
\end{array}
\right].$$
The error parameters $1-p_1$ and $1-p_\phi$ decay exponentially with the idle duration $t$: 
$$1-p_1 = e^{-t/T_1}\quad\textrm{and}\quad1-p_\phi = e^{-t/T_\phi}.$$ 
Note that these channels commute with each other, and also satisfy $$\Lambda(t_1)\circ\Lambda(t_2) = \Lambda(t_1+t_2).$$

In fact, the amplitude- and phase-damping  channels model more than just idling. In a superconducting quantum device, amplitude and phase damping occurs on qubits no matter what operations are being applied. As a result, even though every quantum operation takes a certain duration, many can be modeled as happening instantaneously at a single point of time (usually the middle of its time interval), with the amplitude and phase damping channels applied both before and after. Any operation-specific error channel can also be applied at the same time point. Below, we will focus on the operation-specific errors only.

\subsection{$R_y(\pm\pi/2)$ gates}\label{subsec:Ry}
Following~\cite{obrien2017density}, the gate-specific error of $R_y(\pm\pi/2)$ gates is modeled as depolarizing noise that shrinks the Bloch sphere more in the $x$-$z$ plane and less along the $y$ axis. In the Choi matrix representation:
$$
C_\text{dep}=\left[
\begin{array}{cc|cc}
    1-p_\text{plane}/2 & 0 & 0 & 1-p_\text{plane}/2-p_\text{axis}/2 \\
    0 & p_\text{plane}/2 & -p_\text{plane}/2+p_\text{axis}/2 & 0 \\
    \hline
    0 & -p_\text{plane}/2+p_\text{axis}/2 & p_\text{plane}/2 & 0 \\
    1-p_\text{plane}/2-p_\text{axis}/2 & 0 & 0 & 1-p_\text{plane}/2
\end{array}
\right].$$

\subsection{CZ gates}\label{subsec:cz}
In~\cite{obrien2017density}, the gate-specific error of CZ gates is modeled as a single-qubit phase error (``flux noise") that only applies to the qubit with higher frequency (whose frequency must be shifted away from its ``sweet spot'' to enable the interaction), along with a two-qubit phase error due to flux noise.

Other than single-qubit errors as the qubits are idling, additional errors during the CZ gate are ``quasi-static", meaning that for each pair of coupled qubits, the phase fluctuations $\delta\phi$ and $\delta\phi_\text{2Q}$ are considered constant throughout one run of the circuit, but vary randomly over different runs.

In this article, we choose not to include the quasi-static flux noise mentioned in the Appendix D and E in~\cite{obrien2017density}, because such low-frequency noises can be compensated by rapid qubit calibrations or spin-echo-like noise canceling techniques (e.g. \cite{rol2019fast, debroy2018stabilizer}).

\subsection{Measurement error}\label{subsec:measurement}
The measurement of a qubit is implemented in two steps. In the first step, photons are introduced into a readout resonator, dephasing the qubit completely into the computational basis and allowing the measurement result to be read out. In the second step, the qubit is left idling for a period of time, allowing the photons to deplete from the resonator.

Although measurement is modeled as a ``butterfly gate" in \cite{obrien2017density}, they also note that the parameters observed experimentally can be well explained by simply assuming that the measurement happens instantaneously at the halfway point of the measurement duration, and that the result is subject to a classical declaration error $\epsilon_\text{RO}$ that is independent of the outcome.

Another way measurement can cause errors is through leftover photons in the resonator. Even after photon depletion, there will still be a small number of photons remaining in the cavity, causing the qubit to dephase. The mechanism for this dephasing is complicated and is related to the current state of the qubit, but in this particular circuit, its effect in the time interval $[t_1, t_2]$ can be quantified as
$$p_{\phi, \text{photon}} = \exp\left(2\chi\alpha(0)\exp(\kappa(t_m-t_g))\times\left[\frac{e^{-\kappa t}}{4\chi^2+\kappa^2}[-\kappa\sin(2\chi t)-2\chi\cos(2\chi t)]\right]_{t_1-t_g}^{t_2-t_g}\right),$$
where $t_m$ is the start time of the measurement period, $t_g$ is the time it takes to rotate the qubit from a computational basis state to a Hadamard basis state during an $R_y(-\pi/2)$ gate, and $\kappa$ and $\chi$ are constant parameters. The error channel resulting from this dephasing is similar to the phase damping channel $\Lambda_{T_\phi}$ above, with $p_\phi$ replaced by $p_{\phi, \text{photon}}$ \cite{obrien2017density}.

\subsection{Crosstalk}\label{subsec:crosstalk}
We will consider crosstalk effects between neighboring qubits.
To clarify, we will only consider the crosstalk effect caused by stray 2-qubit interactions.
For example, in~\cite{dicarlo2009demonstration}, it is pointed out that at the flux sweet spot, the ZZ coupling strength is between $0.1\%$ and $1\%$ of the strength at the 2-qubit operating point.
We include the following crosstalk Hamiltonian
\begin{equation}
    H_{\text{xtalk}} = k \sum_{(i,j)} Z_i Z_j =k \sum_{(i,j)} (-I+Z_i+Z_j + 4\ket{11}\bra{11}_{ij}),
    \label{eq:xtalk_hamiltonian}
\end{equation}
where $i, j$ correspond to neighboring qubits.
For simplicity, we assume that the crosstalk strength is uniform and time-independent, which may not be the case in real experiments.
In \autoref{fig_full}, we use CPHASE gates to simulate the effects of crosstalk.
These CPHASE gates correspond to the term $ 4k\ket{11}\bra{11}$ in \autoref{eq:xtalk_hamiltonian}.
From \autoref{eq:xtalk_hamiltonian}, we can see that CPHASE and $e^{ik ZZ}$ are equivalent up to some single qubit $Z$-rotations.
Whether one should use CPHASE or the $e^{ik ZZ}$ gate to model simulations mostly depends on how one calibrates the qubit frequencies.

There are two motivations for using a few instantaneous gates to simulate the effect of crosstalk, which is by nature continuous in time.
The first motivation is to turn the task of solving Lindblad equations into tensor network contraction of gate-based quantum circuits, which consumes significantly less computational resources.
Another motivation  more relevant to physical experiments is to measure the extent to which the effects of crosstalk can be compensated for by slightly manipulating the original measurement circuit.
In fact, the circuit in \autoref{fig_full} already takes this compensation into account.
We expand on this discussion in \autoref{subsect_crosstalk_discretize}.

Note that we will only consider crosstalk noise inside the CZ-region in \autoref{fig_full}. This is because the CZ-regions occupy the majority of the measurement rounds.

\subsection{Parameters}
We adopt the same set of parameters as in \cite{obrien2017density}. In addition, we denote the crosstalk strength by $k$, which we newly include into the noise model. For the sake of readability, we summarize these parameters in \autoref{table_param}.
\begin{table}[h!]
\centering
\begin{tikzpicture}\label{parameters}
\node (tbl) {
\begin{tabularx}{.9\textwidth}{c X r c c }
\arrayrulecolor{purple}
\textbf{Category} & \textbf{Parameter} & \textbf{Symbol} & \textbf{Value} \\
\midrule
scheduling / idle error [\ref{subsec:idle}]& Qubit relaxation time & $T_1$ & $30\mu s$ \\
\midrule
scheduling / idle error [\ref{subsec:idle}]& Qubit dephasing time & $T_{\phi}$ & $60\mu s$ \\
\midrule
scheduling / idle error [\ref{subsec:idle}] & Single-qubit gate time & $T_{g,1Q}$ & 20ns \\
\midrule
scheduling / idle error [\ref{subsec:idle}] & Two-qubit gate time & $T_{g,2Q}$ & 40ns \\
\midrule
scheduling / idle error [\ref{subsec:idle}] & Coherent step time & $\tau_c$ & 200ns \\
\midrule
scheduling / idle error [\ref{subsec:idle}] & Depletion time & $\tau_d$ & 300ns \\
\midrule
scheduling / idle error [\ref{subsec:idle}] & Fast depletion time & $\tau_d^{(fast)}$ & 100ns \\
\midrule
scheduling / idle error [\ref{subsec:idle}] & Measurement time & $\tau_m$ & 300ns \\
\midrule
$R_y(\pm\pi/2)$ gates [\ref{subsec:Ry}] & In-axis rotation error & $p_{axis}$ & $10^{-4}$ \\
\midrule
$R_y(\pm\pi/2)$ gates [\ref{subsec:Ry}] & In-plane rotation error & $p_{plane}$ & $5\times 10^{-4}$ \\
\midrule
measurement error [\ref{subsec:measurement}] & Fast measurement time & $\tau_m^{(fast)}$ & 100ns \\
\midrule
measurement error [\ref{subsec:measurement}] & Readout infidelity & $\epsilon_{RO}$ & $0.15\%$ \\
\midrule
measurement error [\ref{subsec:measurement}] & Photon relaxation time & $\frac{1}{\kappa}$ & 250ns \\
\midrule
measurement error [\ref{subsec:measurement}] & Dispersive shift & $\frac{\chi}{\pi}$ & -2.6 MHz\\
\midrule
crosstalk [\ref{subsec:crosstalk}]& Crosstalk strength & $k$ & $0.03 \sim 0.05$ \\
\midrule
\end{tabularx}};
\end{tikzpicture}
\caption{Experimental parameters used in our simulation}
\label{table_param}
\end{table}

\section{Simulation Techniques}
\subsection{Tensor Network Contraction for Partial-amplitude Computation}
AC-QDP features a strong quantum simulator computing one or multiple amplitudes given a gate-based quantum circuit. The system includes a ``translator'' converting gate-based quantum circuits into tensor networks, together with a tensor network library optimized for contraction tasks on a distributed cluster. In previous reports~\cite{CZH+18,ZHN+19, HSZ+19}, we focused on single-amplitude computation, which translates to the contraction of closed tensor networks (i.e. tensor networks with no open indices). In this work, we extend this framework in order to accommodate the task of evaluating a slice of amplitudes,  i.e. the computational task of contracting a tensor network with open edges.

When contracting a tensor network, all closed indices are summed over one by one. Each time an index is summed over, all tensors adjacent to the index will be merged into one single tensor.
 The time and space complexities can be well approximated by the sum and the maximum of the sizes of the intermediate tensors, respectively. For open tensor networks, there are open indices that are not to be summed over. One then needs to find a contraction order of all the closed indices in order to 
reduce the time and space complexities. In our experiment, we use tree-decomposition based methods to determine the contraction order of the closed indices. More details are provided in \autoref{app:tree_decomp}.

Our engine features a dynamic splitting scheme of the contraction task into smaller subtasks to better exploit the distributed nature of our computational resources~\cite{CZH+18,ZHN+19}. In the case of closed tensor networks, the output of each subtask is a scalar, and it suffices to minimize the workload on each cluster node given the total number of the nodes. For open tensor networks, a subtask may output a large tensor instead, incurring a huge communication cost among the cluster nodes. Therefore, one needs to take into consideration both the workload and the communication cost for each cluster node in order to achieve feasible parallelization.

Tensor-network-based contraction algorithms are usually more efficient than the state vector update algorithm. However, finding the optimal contraction order is NP-hard, and even approximating it might take exponential time~\cite{MS08}. In practice, the circuit simulation is divided into two phases, namely the preprocessing phase in which a near-optimal contraction order is found, and the computation phase where actual contraction happens. One advantage of separating the two phases is that the same contraction order can be used for quantum circuits sharing the same tensor network structure. Therefore, once preprocessing for a particular circuit structure is done, one can freely evaluate the circuit under different noise parameters. This allows us to efficiently compare the results obtained under different noise parameter settings.

\subsection{Comparison with Sampling-Based Approaches}\label{subsec:comparison}

There have been several results focusing on simulating surface-17 under physically motivated noise models~\cite{obrien2017density, tomita2014low, darmawan2017tensor}, a majority of which adopt sampling-based simulation techniques~\cite{obrien2017density, tomita2014low}. Here,  by ``sampling-based'' we refer to methods that repeatedly sample from the underlying distribution of the syndrome outcomes and perform simulations conditioned on the particular syndrome.

Although each trial may take a relatively short time, sampling-based approaches usually suffer from the large number of random trials needed in order to reduce the variance and obtain a sufficiently accurate estimation. On the other hand, exact simulation methods, such as tensor-network-based approaches, take significantly longer time for one trial but do not need repeated experiments. It is not clear \textit{a priori} which simulation method one should adopt, but we present here a simple argument that tensor-network-based simulation is more feasible in our regime of interest.

Suppose that each random trial involves a full density-matrix simulation, which takes time exponential with respect to the number of qubits in the system. In our noise model, it suffices to consider at most 13 qubits at the same time, and the cost for running one trial is roughly $2^{26}$ basic arithmetic operations. To achieve an accuracy of $10^{-5}$ for the logical channel entries, one needs about $10^{10}\approx 2^{33}$ trials, resulting in a total running time of $2^{59}$. A tensor-network based approach, however, runs in time exponential with respect to the treewidth of the tensor network. For the same noise model, with 2+1 syndrome extraction rounds, the resulting tensor network has a treewidth upper bounded by 38. The tensor network-based simulation thus takes time $2^{38}$, over one million times more efficient than sampling-based approaches. 

Of course, such comparisons depend heavily on the specifics of the experiment to be conducted, and there could be cases where sampling-based techniques are feasible while exact computation is not. Nevertheless, tensor-network-based simulation allows us to simulate a physically well-motivated model in the presence of crosstalk.

\section{Numerical Simulations}

\subsection{Experiment Specifications}
In order to support our numerical simulation experiments, Alibaba Cloud Quantum Development Platform was deployed using two different methods. Our first deployment uses Alibaba Cloud's ecs.gn5-c4g1.xlarge instances, each of which has 4 vCPUs, 30 gigabytes of memory and one Nvidia Tesla P100 graphics processing unit with 16 gigabytes of GPU memory. Our second deployment uses Alibaba Cloud's Function Compute Service, a fully-managed event-driven compute service featuring real-time auto scaling and dynamic load balancing within milliseconds. We use up to $1000$ concurrent function executions in our experiments.

\subsection{Effects of crosstalk discretization}
\label{subsect_crosstalk_discretize}

\begin{figure}[!hbt]
    \centering
    \begin{subfigure}{0.45\textwidth}
        \includegraphics[width=\textwidth]{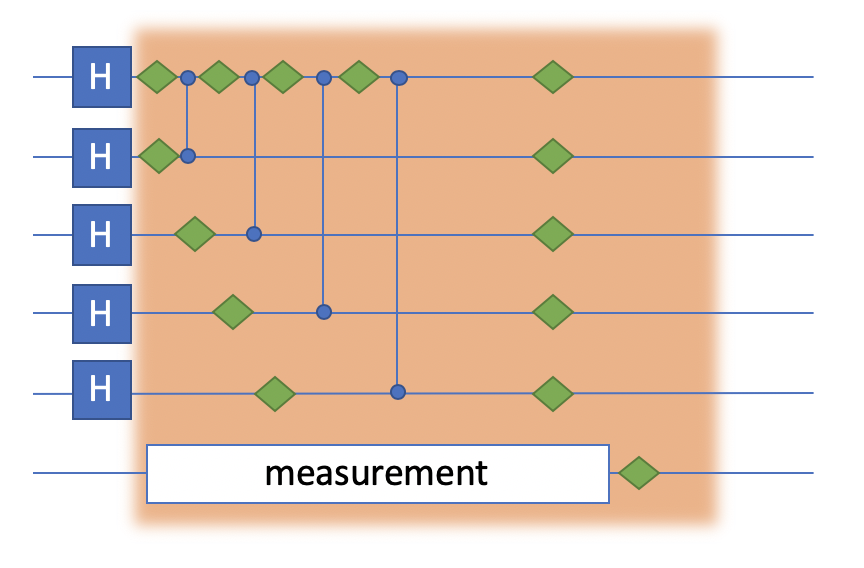}
        \caption{}
    \end{subfigure}
    \begin{subfigure}{0.45\textwidth}
        \includegraphics[width=\textwidth]{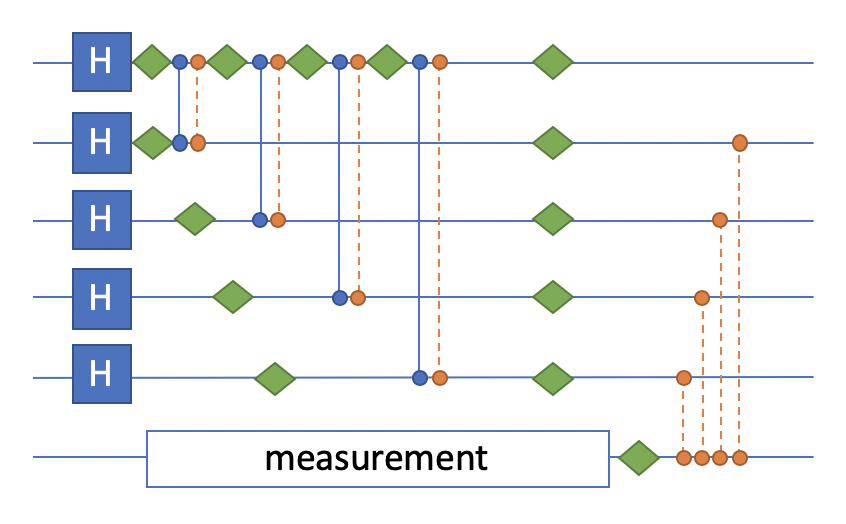}
        \caption{}
    \end{subfigure}
    \newline
    \hspace*{-0.4cm}
    \begin{subfigure}{0.45\textwidth}
        \includegraphics[width=\textwidth]{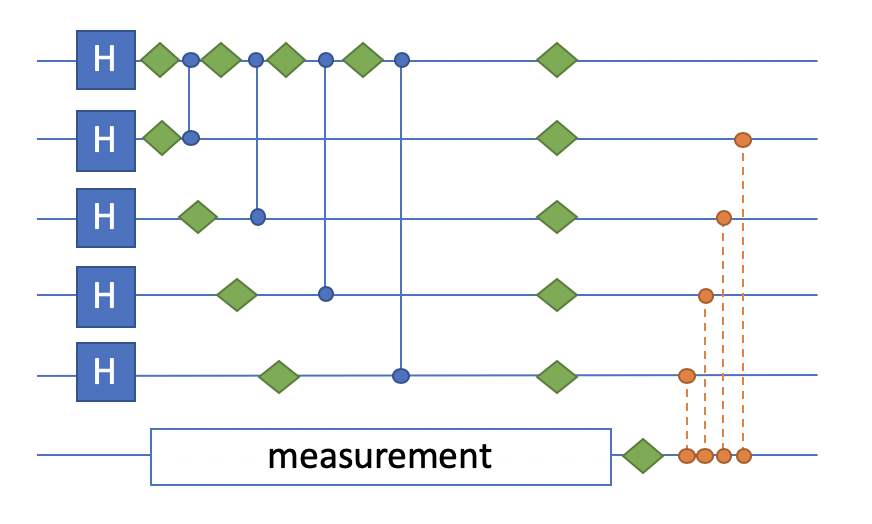}
        \caption{}
    \end{subfigure}\hspace*{0.2cm}
    \begin{subfigure}{0.45\textwidth}
        \includegraphics[width=\textwidth]{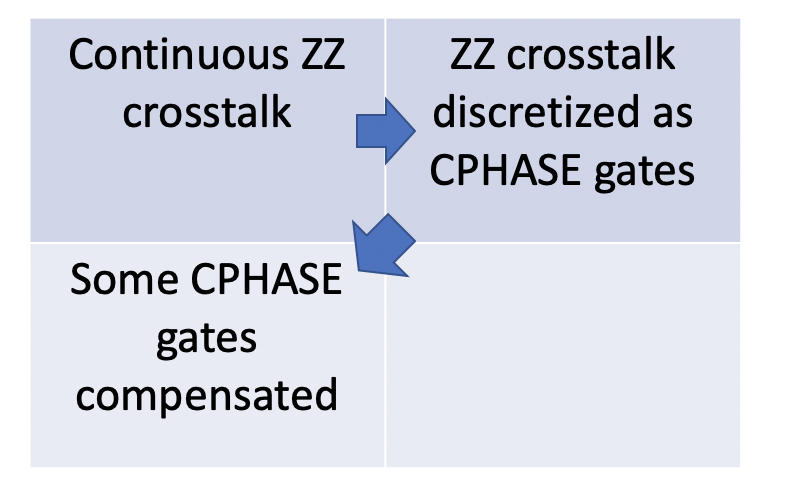}
    \end{subfigure}
    \caption{An illustration of the discretization inaccuracy and the compensation scheme. (a) The continuous crosstalk model. (b) The crosstalk error discretized as instantaneous CPHASE gates at specific positions. The inaccuracy of simulation incurred by going from (a) to (b) is referred to as the discretization inaccuracy. (c) A compensation scheme absorbing the CPHASE gate into CZ rotations.}
    \label{fig_disc}
\end{figure}

As we briefly mentioned in \autoref{subsec:crosstalk}, one goal of the simulation is to determine to what degree we can reasonably discretize the crosstalk noise.
For a continuous-time simulation, we can include a crosstalk Hamiltonian (\autoref{eq:xtalk_hamiltonian}) in the Lindblad equation.
In the discretization, we can add one control-phase (CPHASE) gate per two neighboring qubits for each round of $X(Z)$-stabilizer measurements.
The CPHASE operation is set to be
    $e^{-4ikt\ket{11}\bra{11}}$,
where $k$ is the crosstalk strength in \autoref{eq:xtalk_hamiltonian} and $t$ is the time duration of the CZ-region.
We call the difference between these two simulations ``the discretization inaccuracy''. 
The discretization approach is potentially much easier computationally, especially when using tensor networks for data representation. Furthermore, such discretization allows us to partially compensate for the effect of the crosstalk by adjusting the CZ rotation.
Therefore, we want to determine how large the discretization inaccuracies are. See \autoref{fig_disc} for an illustration of the discretization inaccuracy and the compensation scheme.

\begin{figure}
    \centering
    \includegraphics[width=\linewidth]{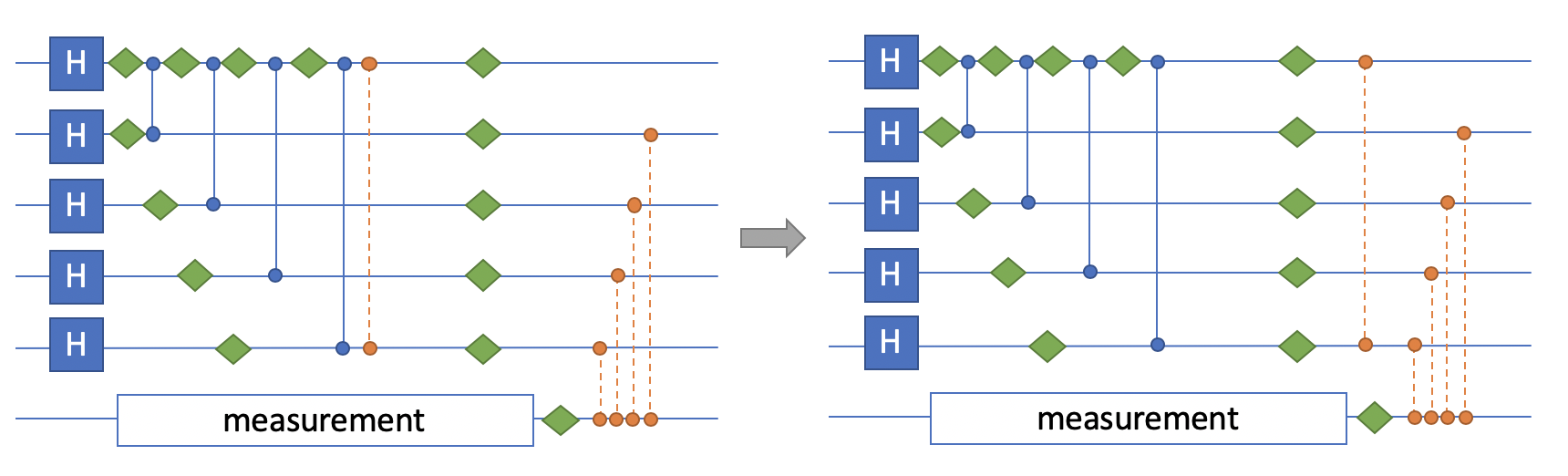} 
    \caption{One particular way of moving a CPHASE-gate.}\label{fig_moving_error}
\end{figure}

However, a direct computation of the discretization inaccuracy may require solving the Lindblad equation, which is beyond our computational power. Instead, we estimate the discretization inaccuracy using the so-called ``moving inaccuracy'', i.e., the amount of inaccuracy created by moving $\text{CPHASE}(\Delta \theta)$ inside the CZ-region (see \autoref{fig_moving_error}).

It is easy to see how the moving inaccuracy and discretization inaccuracy are related.
If we can move $\text{CPHASE}(\Delta \theta)$ arbitrarily inside a CZ-region while only causing very small change to the whole channel, then we can move all crosstalk on a pair of qubits to the same location.
Because $\text{CPHASE}(\Delta \theta)$ is diagonal, we can commute it through CZ gates, but not the amplitude-phase damping.
Therefore, the moving inaccuracy is only generated from interchanging $\text{CPHASE}(\Delta \theta)$ and amplitude-phase damping.

If the discretization/moving inaccuracies are low, an immediate consequence is that we can compensate for the crosstalk errors inside the CZ-region by adjusting the conditional rotation angles of CZ-gates. We can also obtain an indication of how large the inaccuracy is by not interpolating amplitude damping errors during CZ gates.

In principle, the discretization inaccuracy and the moving inaccuracy can be upper-bounded by the commutator norm of the discretized crosstalk term $\text{CPHASE}(\Delta \theta)$ and the error channel within the CZ-region. However, direct calculation of the commutator yields a norm on the magnitude of $10^{-3}$, too large in the context of simulating error-correction circuits. Surprisingly, this can be mitigated because most components of such inaccuracies result in correctable errors for the error-correction circuit, and the residual inaccuracy is orders of magnitude smaller than the commutator norm, as we show below using numerical experiments.

In the numerical simulation, we estimate the moving inaccuracy by considering the particular error induced by moving the ZZ crosstalk term on the neighboring qubits (3,3) and (2,4) within the CZ-region. For one round of noisy syndrome extraction, we compare the logical channel with and without moving CPHASE gates in \autoref{table_moving_error}.  We observe that the moving inaccuracy is on the order of $10^{-8}$, and thus is inconsequential for the simulation of error-correction in the current noise regime.

\begin{table}[bht!]
    \centering
    \begin{subtable}{.7\linewidth}\centering
        \begin{tikzpicture}[every node/.style={anchor=base,node distance=1cm, text depth=0.5ex,text height=1.5em,text width=4em}]
            \matrix [inner sep=2mm]
            {
            \node[shape=rectangle, fill={rgb:black,1;white,2},inner sep=2.8mm] {1.00e+00}; & \node[fill={rgb:black,1;white,2},inner sep=2.8mm] {-1.22e-17}; & \node[fill={rgb:black,1;white,2},inner sep=2.8mm] {-1.57e-21}; & \node[fill={rgb:black,1;white,2},inner sep=2.8mm] {8.56e-17}; \\
            \node[fill=blue!30!white,inner sep=2.8mm] {9.90e-08}; & \node[fill=white!90!green,inner sep=2.8mm] {9.99e-01}; & \node[fill=yellow!30!white,inner sep=2.8mm,inner sep=2.8mm] {-3.24e-06}; & \node[fill=yellow!30!white,inner sep=2.8mm] {-2.19e-17}; \\
            \node[fill=blue!30!white,inner sep=2.8mm] {-5.52e-11}; & \node[fill=yellow!30!white,inner sep=2.8mm] {3.17e-06}; & \node[fill=white!90!green,inner sep=2.8mm,inner sep=2.8mm] {9.98e-01}; & \node[fill=yellow!30!white,inner sep=2.8mm] {1.52e-21}; \\
            \node[fill=blue!30!white,inner sep=2.8mm] {9.50e-08}; & \node[fill=yellow!30!white,inner sep=2.8mm] {-1.24e-11}; & \node[fill=yellow!30!white,inner sep=2.8mm,inner sep=2.8mm] {-6.29e-21}; & \node[fill=white!90!green,inner sep=2.8mm] {9.99e-01}; \\
            };
            \end{tikzpicture}
        \subcaption{Logical PTM without moving the error}\label{tab:1a}
    \end{subtable}
    \newline
    \vspace*{0.1cm}    
    \newline    \begin{subtable}{.7\linewidth}\centering
        \begin{tikzpicture}[every node/.style={anchor=base,node distance=1cm, text depth=0.5ex,text height=1.5em,text width=4em}]
            \matrix [inner sep=2mm]
            {
            \node[shape=rectangle, fill={rgb:black,1;white,2},inner sep=2.8mm] {1.00e+00}; & \node[fill={rgb:black,1;white,2},inner sep=2.8mm] {-1.49e-17}; & \node[fill={rgb:black,1;white,2},inner sep=2.8mm] {-2.90e-21}; & \node[fill={rgb:black,1;white,2},inner sep=2.8mm] {-3.31e-17}; \\
            \node[fill=blue!30!white,inner sep=2.8mm] {9.90e-08}; & \node[fill=white!90!green,inner sep=2.8mm] {9.99e-01}; & \node[fill=yellow!30!white,inner sep=2.8mm,inner sep=2.8mm] {-3.24e-06}; & \node[fill=yellow!30!white,inner sep=2.8mm] {-4.91e-14}; \\
            \node[fill=blue!30!white,inner sep=2.8mm] {-5.52e-11}; & \node[fill=yellow!30!white,inner sep=2.8mm] {3.17e-06}; & \node[fill=white!90!green,inner sep=2.8mm,inner sep=2.8mm] {9.98e-01}; & \node[fill=yellow!30!white,inner sep=2.8mm] {6.04e-10}; \\
            \node[fill=blue!30!white,inner sep=2.8mm] {9.50e-08}; & \node[fill=yellow!30!white,inner sep=2.8mm] {-1.24e-11}; & \node[fill=yellow!30!white,inner sep=2.8mm,inner sep=2.8mm] {-6.11e-10}; & \node[fill=white!90!green,inner sep=2.8mm] {9.99e-01}; \\
            };
            \end{tikzpicture}
        \subcaption{Logical PTM with the error moved}\label{tab:1b}
    \end{subtable}
    \newline
    \vspace*{0.1cm}    
    \newline
    \begin{subtable}{.7\linewidth}\centering
        \begin{tikzpicture}[every node/.style={anchor=base,node distance=1cm, text depth=0.5ex,text height=1.5em,text width=4em}]
            \matrix [inner sep=2mm]
            {
            \node[shape=rectangle, fill={rgb:black,1;white,2},inner sep=2.8mm] {8.88e-16}; & \node[fill={rgb:black,1;white,2},inner sep=2.8mm] {-2.67e-18}; & \node[fill={rgb:black,1;white,2},inner sep=2.8mm] {-1.33e-21}; & \node[fill={rgb:black,1;white,2},inner sep=2.8mm] {-1.19e-16}; \\
            \node[fill=blue!30!white,inner sep=2.8mm] {1.18e-13}; & \node[fill=white!90!green,inner sep=2.8mm] {-8.57e-09}; & \node[fill=yellow!30!white,inner sep=2.8mm,inner sep=2.8mm] {-1.90e-12}; & \node[fill=yellow!30!white,inner sep=2.8mm] {-4.91e-14}; \\
            \node[fill=blue!30!white,inner sep=2.8mm] {9.01e-17}; & \node[fill=yellow!30!white,inner sep=2.8mm] {1.01e-12}; & \node[fill=white!90!green,inner sep=2.8mm,inner sep=2.8mm] {-2.21e-08}; & \node[fill=yellow!30!white,inner sep=2.8mm] {6.04e-10}; \\
            \node[fill=blue!30!white,inner sep=2.8mm] {4.15e-13}; & \node[fill=yellow!30!white,inner sep=2.8mm] {-3.43e-16}; & \node[fill=yellow!30!white,inner sep=2.8mm,inner sep=2.8mm] {-6.11e-10}; & \node[fill=white!90!green,inner sep=2.8mm] {-1.53e-08}; \\
            };
            \end{tikzpicture}
        \subcaption{The difference incurred by moving the error term}\label{tab:1c}
    \end{subtable}
    \newline
    \caption{Comparison of logical channels with and without moving the crosstalk term, and the difference between the two. We choose to include the exact result of the first row to illustrate the magnitude of numerical error.}
    \label{table_moving_error}
\end{table}

To study the moving inaccuracy under different noise parameter regimes, we do experiments to determine whether the difference between the two channels, quantified by the $1$-norm of the PTM, changes with respect to a change in any single parameter in the noise model. The results are shown in \autoref{table_projected}, where various parameters are perturbed, and the two moving inaccuracies corresponding to the qubit pairs (3,3)-(2,4) and (1,3)-(0,2) are considered. It shows that moving the crosstalk term incurs only a very small amount of inaccuracy for noise model parameters in our regime of interest, and thus a large portion of the crosstalk can be mitigated by phase compensation on the CPHASE gates.

\begin{table}[bth!]
    \begin{tabular}{c|cc}
        parameter changed & moving inaccuracy on (3,3)-(2,4)& moving inaccuracy on (1,3)-(0,2)\\
        \hline
        unchanged & 7.36e-7&1.59e-7\\
        \hline
        $T_1\leftarrow 3\mu s$ &3.51e-5&1.24e-5\\
        $T_1\leftarrow 10\mu s$ &5.24e-6&1.59e-6\\
        \hline
        $T_\phi\leftarrow 6\mu s$ &4.35e-7&4.44e-7\\
        $T_\phi\leftarrow 20\mu s$ &6.20e-7&2.08e-7\\
        \hline
    \end{tabular}
    \caption{The change of the moving inaccuracy with different noise parameters $T_1$ and $T_\phi$ }
    \label{table_projected}
\end{table}

\subsection{Effects of crosstalk for 2+1-round syndrome extraction}
To compute the logical PTM under the optimal decoder, one needs to compute all the PTMs corresponding to quantum operations for each assignment of the syndrome bits. AC-QDP can handle up to 2+1 rounds of syndrome extraction where a total of 24 syndrome bits are extracted. Note that this is the depth proposed for a near-term fault-tolerance demonstration in~\cite{trout2018simulating}.

We report the logical channels for $2+1$ rounds of syndrome extraction, with and without the presence of the crosstalk, shown in \autoref{table_2c}.
\begin{table}[ht!]
\centering
\begin{subtable}{.7\linewidth}\centering
    \begin{tikzpicture}[every node/.style={anchor=base,node distance=1cm, text depth=0.5ex,text height=1.5em,text width=4em}]
        \matrix [inner sep=2mm]
        {
        \node[shape=rectangle, fill={rgb:black,1;white,2},inner sep=2.8mm] {1.00e+00}; & \node[fill={rgb:black,1;white,2},inner sep=2.8mm] {-4.07e-18}; & \node[fill={rgb:black,1;white,2},inner sep=2.8mm] {0.00e+00}; & \node[fill={rgb:black,1;white,2},inner sep=2.8mm] {-9.42e-18}; \\
        \node[fill=blue!30!white,inner sep=2.8mm] {2.88e-07}; & \node[fill=white!90!green,inner sep=2.8mm] {9.92e-01}; & \node[fill=yellow!30!white,inner sep=2.8mm,inner sep=2.8mm] {0.00e+00}; & \node[fill=yellow!30!white,inner sep=2.8mm] {-2.29e-13}; \\
        \node[fill=blue!30!white,inner sep=2.8mm] {0.00e+00}; & \node[fill=yellow!30!white,inner sep=2.8mm] {0.00e+00}; & \node[fill=white!90!green,inner sep=2.8mm,inner sep=2.8mm] {9.85e-01}; & \node[fill=yellow!30!white,inner sep=2.8mm] {0.00e+00}; \\
        \node[fill=blue!30!white,inner sep=2.8mm] {6.14e-06}; & \node[fill=yellow!30!white,inner sep=2.8mm] {-2.31e-13}; & \node[fill=yellow!30!white,inner sep=2.8mm,inner sep=2.8mm] {0.00e+00}; & \node[fill=white!90!green,inner sep=2.8mm] {9.92e-01}; \\
        };
        \end{tikzpicture}
    \subcaption{Logical PTM for 2+1 rounds of syndrome extraction without crosstalk}\label{tab:2a}
\end{subtable}
\newline
\vspace*{0.1cm}    
\newline
\begin{subtable}{.7\linewidth}\centering
    \begin{tikzpicture}[every node/.style={anchor=base,node distance=1cm, text depth=0.5ex,text height=1.5em,text width=4em}]
        \matrix [inner sep=2mm]
        {
        \node[shape=rectangle, fill={rgb:black,1;white,2},inner sep=2.8mm] {1.00e+00}; & \node[fill={rgb:black,1;white,2},inner sep=2.8mm] {-2.44e-18}; & \node[fill={rgb:black,1;white,2},inner sep=2.8mm] {-5.37e-20}; & \node[fill={rgb:black,1;white,2},inner sep=2.8mm] {5.96e-18}; \\
        \node[fill=blue!30!white,inner sep=2.8mm] {3.47e-07}; & \node[fill=white!90!green,inner sep=2.8mm] {9.88e-01}; & \node[fill=yellow!30!white,inner sep=2.8mm,inner sep=2.8mm] {-2.65e-03}; & \node[fill=yellow!30!white,inner sep=2.8mm] {2.99e-08}; \\
        \node[fill=blue!30!white,inner sep=2.8mm] {4.12e-09}; & \node[fill=yellow!30!white,inner sep=2.8mm] {-2.63e-03}; & \node[fill=white!90!green,inner sep=2.8mm,inner sep=2.8mm] {9.80e-01}; & \node[fill=yellow!30!white,inner sep=2.8mm] {-7.43e-06}; \\
        \node[fill=blue!30!white,inner sep=2.8mm] {8.89e-06}; & \node[fill=yellow!30!white,inner sep=2.8mm] {-6.75e-10}; & \node[fill=yellow!30!white,inner sep=2.8mm,inner sep=2.8mm] {7.95e-06}; & \node[fill=white!90!green,inner sep=2.8mm] {9.91e-01}; \\
        };
        \end{tikzpicture}
    \subcaption{Logical PTM for 2+1 rounds of syndrome extraction with crosstalk}\label{tab:2b}
\end{subtable}
\newline
\vspace*{0.1cm}    
\newline
\begin{subtable}{.7\linewidth}\centering
        \begin{tikzpicture}[every node/.style={anchor=base,node distance=1cm, text depth=0.5ex,text height=1.5em,text width=4em}]
            \matrix [inner sep=2mm]
            {
            \node[shape=rectangle, fill={rgb:black,1;white,2},inner sep=2.8mm] {0.00e+00}; & \node[fill={rgb:black,1;white,2},inner sep=2.8mm] {-1.62e-18}; & \node[fill={rgb:black,1;white,2},inner sep=2.8mm] {5.37e-20}; & \node[fill={rgb:black,1;white,2},inner sep=2.8mm] {-1.54e-17}; \\
            \node[fill=blue!30!white,inner sep=2.8mm] {-5.86e-08}; & \node[fill=white!90!green,inner sep=2.8mm] {4.43e-03}; & \node[fill=yellow!30!white,inner sep=2.8mm,inner sep=2.8mm] {2.65e-03}; & \node[fill=yellow!30!white,inner sep=2.8mm] {-2.98e-08}; \\
            \node[fill=blue!30!white,inner sep=2.8mm] {-4.12e-09}; & \node[fill=yellow!30!white,inner sep=2.8mm] {2.63e-03}; & \node[fill=white!90!green,inner sep=2.8mm,inner sep=2.8mm] {4.84e-03}; & \node[fill=yellow!30!white,inner sep=2.8mm] {7.43e-06}; \\
            \node[fill=blue!30!white,inner sep=2.8mm] {-2.74e-06}; & \node[fill=yellow!30!white,inner sep=2.8mm] {6.75e-10}; & \node[fill=yellow!30!white,inner sep=2.8mm,inner sep=2.8mm] {-7.95e-06}; & \node[fill=white!90!green,inner sep=2.8mm] {6.54e-04}; \\
            };
            \end{tikzpicture}
        \subcaption{The difference between the two logical PTMs}\label{tab:2c}
\end{subtable}
\newline
\caption{Comparison of logical channels with and without crosstalk for 2+1 rounds of syndrome extraction}
\label{table_2c}
\end{table}
From the table, we can infer that the effect of ZZ crosstalk on the logical channel is concentrated on the logical coherent Z rotation and the stochastic phase-flip error, each of  magnitude $\sim 10^{-3}$.

\begin{figure}[hbtp]
    \centering
    \includegraphics[width=0.7\textwidth]{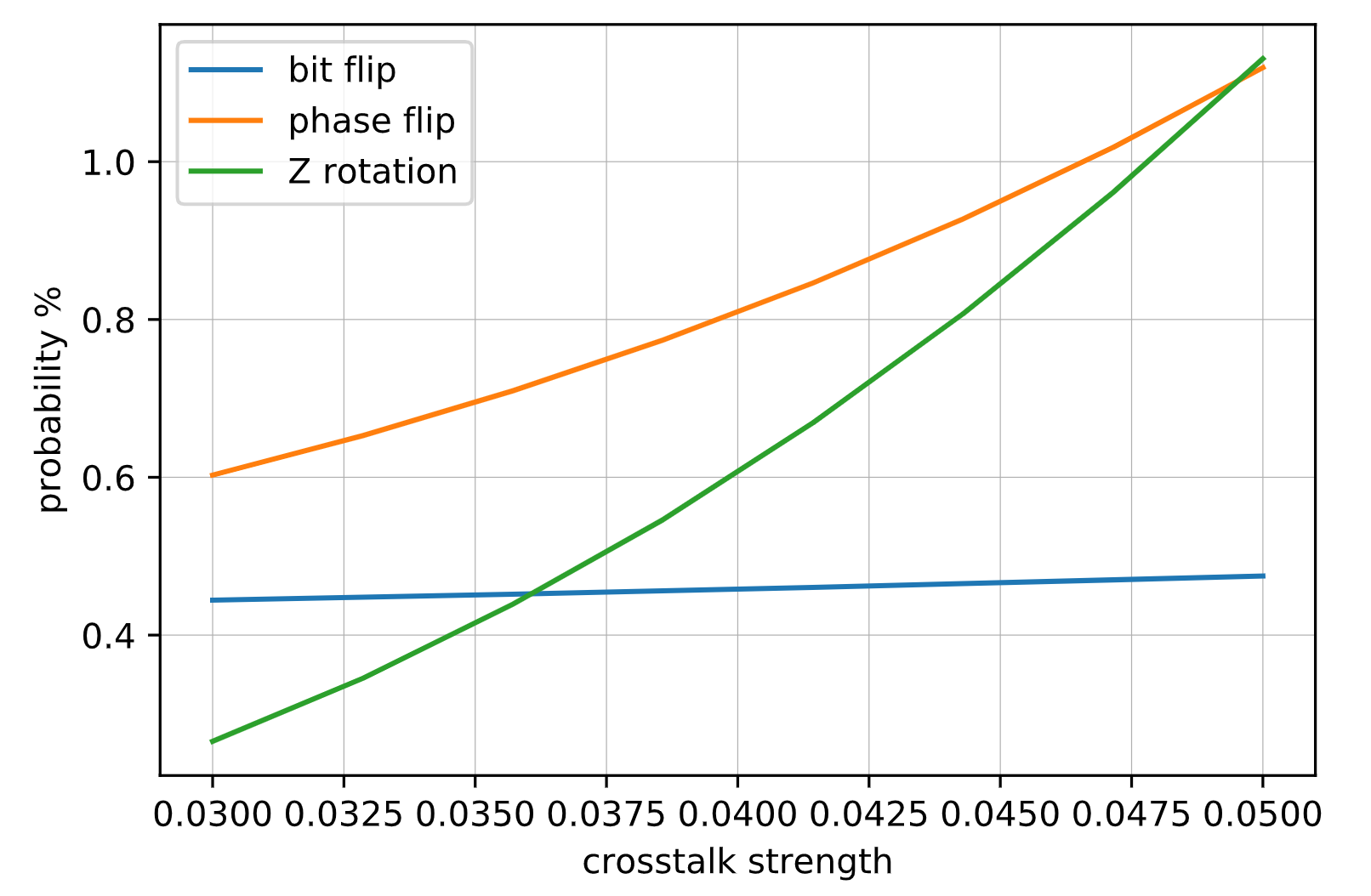}
    \caption{Bit-flip probability and phase-flip probability}
    \label{fig_sensitivity}
\end{figure}

To study the sensitivity of bit-flip, phase-flip and coherent Z rotation errors with respect to the crosstalk strength, we performed simulations on varying crosstalk strengths from $0.3$ to $0.5$ while keeping all other parameters fixed. It can be seen from \autoref{fig_sensitivity} that the phase-flip error rate is much more sensitive to crosstalk than the bit-flip error rate. When the crosstalk strength is $0.5$, the probability that a logical phase error occurs exceeds $1\%$. The relative sensitivity of phase-flip error to crosstalk is likely due to the fact that X-type errors on the ancilla qubits, enhanced by the photon decay effect, are more likely to introduce hook errors through the crosstalk channels and result in logical phase-flips. Z errors on the ancilla qubits, on the other hand, commute with ZZ crosstalk terms and do not result in damaging errors.

\section{Summary and Outlook}
In this article, we performed large-scale simulations of a logical memory experiment using the surface code with $17$ qubits. In contrast to previous works \cite{obrien2017density, tomita2014low, darmawan2017tensor}, our tensor-network-based simulation is able to handle ZZ-interaction crosstalk on neighboring pairs of qubits.  Crosstalk remains a significant experimental error, with physical solutions like tunable couplers \cite{arute2019quantum} or sparse lattice geometries \cite{chamberland2020topological, castro2019generating, chamberland2020triangular} being proposed to combat it.  Consequently, it is necessary to consider crosstalk in order to accurately assess the performance of near-term error-correction experiments.

We highlight the advantage of performing tensor-network-based simulation with AC-QDP. As was noted in \autoref{subsec:comparison}, a tensor-network-based approach can be more efficient than sampling-based approaches in order to achieve a certain accuracy for the logical channel. In addition, though the simulation results we present only require a single GPU machine or hundreds of concurrencies through the FunctionCompute service, AC-QDP will be capable of handling larger tensor network contractions by dynamically seeking an optimal strategy for splitting a large task and then distributing the subtasks through the cloud.

It is important to note that we only considered crosstalk interactions between neighboring qubits. Although these ZZ-interactions only yield a second-order effect on the logical channel, as such errors are fault-tolerantly accounted for within the standard depolarizing model, our simulation shows that they nonetheless contribute significantly to the logical dephasing. This indicates that for the code considered, the current noise parameters approach, but do not yet meet, the break-even fault-tolerance regime, even when ignoring potentially uncorrectable correlated noise.
\subsection{Future Work}

\paragraph{Crosstalk.}
In this work, we only considered the effects of ZZ-type crosstalk on the logical channel. Such errors only constitute a small fraction of unwanted coupling between qubits. For example, there is electromagnetic crosstalk such as qubit control crosstalk~\cite{BAH+11} and measurement crosstalk~\cite{ACS+10} between neighboring qubits and other stray interactions in the Hamiltonian~\cite{SMC+16}. We leave it to future work to perform device-specific simulations to account for these various forms of crosstalk.

\paragraph{Reset.}
The reset procedure is used to prepare the ground state $|0\rangle$ with high fidelity in the middle of a quantum circuit. Reset schemes are usually much faster than the natural energy depletion process characterized by the amplitude damping time $T_1$.  In this article, we chose not to include the reset scheme, as the reset and the measurement time may be a significant bottleneck in current superconducting experiments.  Additionally, a scheme similar to reset, called feedback control, was implemented in~\cite{obrien2017density} to determine if error rates could be further suppressed. The feedback control scheme applies an X gate to the ancilla qubit at the end of a measurement round if and only if the measurement result in the previous round is $-1$. Such a scheme keeps an ancilla qubit at $|0\rangle$ between rounds unless some error occurs during the most recent round which flips the corresponding syndrome bit. Assuming that errors happen infrequently, the feedback scheme desensitizes ancilla qubits to amplitude damping. However, the experiment using the feedback scheme showed only a small improvement, providing evidence that even perfect reset schemes might not be of significant help in the current noise regime. 

It is worth noting that the state-of-the-art reset schemes achieve fidelities in excess of $99.8\%$ within $500$ns~\cite{magnard2018fast}, which could be fast enough for use within an error-correction scheduling in superconducting systems. We leave it to future work to simulate an error-correction scheme with reset incorporated.

\paragraph{Leakage.}
Throughout the simulation, we assumed that all qubits stay in the lowest-energy two-level system, ignoring the possibility that a qubit state may leak into a higher energy subspace. Although the probability of leakage can be small in superconducting systems~\cite{wood2018quantification}, leakage errors can be devastating, as they are not correctable by standard error-correction techniques. Fortunately, several schemes at both the physical and circuit level have been proposed in order to suppress leakage errors \cite{terhal2019leakage, suchara2015leakage, aliferis2005fault, fowler2013coping, ghosh2013understanding, ghosh2015leakage, chen2016measuring, brown2019handling, ghosh2017pulse}.

In this work, we chose not to include leakage as a leaked system would have to be modeled by (at least) a three-level system.  This increases the bond dimension of some of the hyperedges in the tensor network from 2 to 3, and consequently increases the computational cost beyond our computational power.  
In addition, leakage errors require additional circuitry or control to eliminate. Owing to their low probability, they may not prove a limiting factor in low-depth near-term experiments. Therefore, we leave the simulation and correction of physically-motivated leakage models to future work.

Recently, building on simulation tricks in~\cite{obrien2017density}, numerical experiments were performed in limited but physically-motivated leakage models that allow for state compression~\cite{barbara2020leakage}.  While these compressions are unavailable for crosstalk simulation, it would be interesting to try to simultaneously capture both noise processes at a simulable size using a combination of methods.  We leave this to future work.

\section*{Acknowledgements}
We would like to thank our colleagues from various teams within Alibaba Cloud Intelligence supporting us in the numerical experiments presented in this paper.  F.~Z. was supported in part by the US NSF under award 1717523.

\bibliographystyle{plain}
\bibliography{main}

\newpage
\appendix
\begin{section}{Open Tensor Network Contraction}
    \label{app:tree_decomp}
    \subsection{Tensor network and contraction}
    In the context of the article, it suffices to consider a tensor as a multi-index array over complex numbers. For sake of simplicity, we assume that all the indices run over $\{0, 1\}$. The number of indices attached to a tensor is called the rank of the tensor.

    A tensor network $N$ can be regarded as a multi-hypergraph $N=(V,E)$, where each node $v\in V$ is associated to a tensor $T_v$, and each hyperedge $e$ corresponds to an index connecting all the adjacent tensors. The contraction of a hyperedge $e\in E$ is to remove $e$ from the tensor network, and to replace all its neighboring tensors by a single tensor obtained by summing over the values of their shared index $e$. The value of a tensor network $N$, denoted by $val(N)$, is the final scalar obtained from contracting all of the indices in an arbitrary order. In the case of open tensor networks, there is a subset $E_o\subset E$ of indices that are not to be contracted. The final result would then be a tensor of rank $|E_o|$ obtained by contracting all of the closed (non-open) edges in an arbitrary order.
   
    Given the tensor network, the computational task of solving the final output tensor is called the contraction of the tensor network. Note that the final result of the tensor network does not depend on the order of contraction; however the time and space complexities vary greatly over different contraction orders. One therefore needs to carefully choose an efficient ordering of the closed indices for contraction.
 
    \subsection{Treewidth and contraction order}
    The time and space complexities associated to a specific contraction order can be well approximated by the maximum size of the intermediate tensors. It is shown in~\cite{MS08} that this quantity can be lower bounded by the treewidth of the line graph of the tensor network. For a tensor network hypergraph $N=(V,E)$, its line graph can be defined as $N'=(E, \{\{e|v\in e, e\in E\}|v\in V\})$, i.e. each index of the tensor network corresponds to a node in the linegraph, and each tensor corresponds to a hyperedge of the line graph.
    
    We first recall the tree decomposition of a (hyper)graph. 

\begin{definition}[Tree decomposition]
    A tree decomposition of a (hyper)graph $G=(V_G,E_G)$ is a tree $T=(V_T,E_T)$, where each node $w$ in $V_T$ is associated to a subset $S_w$ of $V$. Moreover, the tree $T$ needs to satisfy the following properties:
    \begin{itemize}
        \item For all $u\in V$, $\{w\in V_T|u\in S_w\}$ forms a connected subtree of $T$;
        \item For all $e\in E$, there exists $w\in V_T$ such that $e\subset S_w$.
    \end{itemize}
\end{definition}
Each of the subsets $S_w$ is called a bag, and the maximum size over all the bags is called the width of the tree decomposition $T$, denoted as $w(T)$. In the context of tensor networks, each bag of the tree decomposition is a collection of indices appearing in a tensor network.

From the tree decomposition $T$, one can retrieve a contraction order for the tensor network:
\begin{itemize}
    \item Select an arbitrary tree node to be the root of the tree and initialize the contraction order to be empty.
    \item While there is a leaf node in the tree, remove that node. Attach all graph nodes only appearing in the leaf node to the end of the contraction order in an arbitrary order.
\end{itemize}

One can verify that the largest intermediate tensor of such a contraction order is at most exponential in the width of the tree decomposition. Thus, the task of finding a good contraction order can be translated into finding a tree decomposition with as small as possible width. The minimum width of all the tree decompositions of a hypergraph is called the treewidth of the hypergraph.

In the case of open tensor networks, one can simply add a virtual tensor adjacent to all the open indices. This would result in a tree node in the tree decomposition, where there exists a bag that contains all of the open indices. One can recover a contraction order by taking this bag as the root node and only attaching closed indices to the contraction order during the process. One can verify that the maximum size of all the intermediate tensors (including the final output) is again exponential in the width of the tree decomposition. The task of finding a good contraction order then again amounts to a good tree decomposition routine for the augmented tensor network.

\subsection{Splitting of tensor networks}
Similar to~\cite{CZH+18,ZHN+19,HSZ+19}, our software features a dynamic splitting scheme of a contraction task into various smaller subtasks. This is done by enumerating over all possible assignments for a number of selected indices. 

For a rank-$k$ tensor $T_{ab\dots cd}$, fixing an index, e.g. $a$, to a specific value, e.g. $0$, results in a tensor $T'_{b\dots cd}$ of rank $k-1$, such that $T'_{b\dots cd}=T_{0b\dots cd}$. Fixing an index $e$ in a tensor network $N$ to a specific value $v$ removes the index from the tensor network, and fixes it to $v$ on all its adjacent tensors. Note that this results in another tensor network $N_{e:v}$.

When $e$ is a closed edge, we have that $val(N)=\sum_v val(N_{e:v})$. When $e$ is open, the value $val(N)$ is obtained by stacking the tensors $val(N_{e:v})$ along the $e$ direction. In either case, the value $val(N)$ can be obtained from the values of the subtasks. Note that whichever value an edge is fixed to, the sub-tensor networks share the same hypergraph and take the same amount of time to contract. Such a scheme can be easily generalized to splitting multiple edges. The number of subtasks would scale exponentially with the number of split edges, but each sub tensor network can be contracted using less time and space compared to the full task. Given some constraints on the computational cluster (e.g. concurrency, space limit, communication cost), one can then find a selection of edges to split so that the contraction task of the subtasks is feasible and as efficient as possible.

\end{section}

\begin{section}{Discussion on the 2-qubit Pauli errors in the measurement circuit}

In this section, we discuss the effect of 2-qubit Pauli errors corresponding to ZZ-crosstalk in the measurement circuit, as they provide a rough but analyzable perspective to study the 2-qubit components of crosstalk.
The 2-qubit Pauli errors we need to consider are $Z\otimes Z$ errors on neighboring data and ancilla qubits.
It is not hard to show that such a $Z\otimes Z$ error inside a CZ-region in \autoref{fig_full} can be corrected.
This is because ZZ errors commute with the CZ-gates, and we can move it to the time step after the CZ-gate which is then applied on the same two qubits.
Afterwards, we can use the usual fault-tolerance arguments for surface-17 to show that such error can be corrected.
By the above argument, we know that if the crosstalk errors are $e^{i\theta ZZ}$, and if they only happen inside the CZ-regions, then the logical error rate will only contain terms of order $o(\theta)$.

However, a general two-qubit Pauli error occurring on neighboring qubits inside a CZ-region can break fault-tolerance.
One can create such examples by mimicking so called ``hook'' errors~\cite{tomita2014low}.
Similarly, in an alternative syndrome measurement circuit which uses CNOT as its native two qubit gate, a single two-qubit $Z\otimes Z$ error at certain locations can break fault-tolerance.
Therefore, we should keep in mind that the impact of crosstalk on error-correction will heavily depend on the details of crosstalk errors and measurement circuits.

One important thing which we do not cover in the above analysis is the crosstalk noise happening outside the CZ-regions.
In general, crosstalk will occur during the whole time duration, and even during single qubit rotations in \autoref{fig_full}. This means that although the crosstalk Hamiltonian a ZZ-interactions, the corresponding Pauli errors will not only contain ZZ terms. Therefore, at distance-$3$, it is likely that the logical error rate will have some linear dependence on the crosstalk strength.
Another difference between Pauli errors and crosstalk noise is that the crosstalk noise is coherent.
With ZZ crosstalk inside the CZ-regions and without any other error sources, we can simply compensate for crosstalk by implementing additional CPHASE gates or changing the CZ-gates accordingly.
For example, in a CZ-region in which the $Z$-type stabilizers are measured, we can compensate for crosstalk between data qubits and $Z$-ancilla qubits by changing the CZ-gates between them.
By comparison, there is no analogous mechanism to compensate for stochastic Pauli errors.
\end{section}

\clearpage

\begin{section}{Logical PTMs for various simulation settings}

\begin{table}[ht!]
    \centering
    \begin{subtable}{.7\linewidth}\centering
        \begin{tikzpicture}[every node/.style={anchor=base,node distance=1cm, text depth=0.5ex,text height=1.5em,text width=4em}]
            \matrix [inner sep=2mm]
            {
            \node[shape=rectangle, fill={rgb:black,1;white,2},inner sep=2.8mm] {1.00e+00}; & \node[fill={rgb:black,1;white,2},inner sep=2.8mm] {-1.23e-17}; & \node[fill={rgb:black,1;white,2},inner sep=2.8mm] {0.00e+00}; & \node[fill={rgb:black,1;white,2},inner sep=2.8mm] {8.61e-17}; \\
            \node[fill=blue!30!white,inner sep=2.8mm] {1.29e-07}; & \node[fill=white!90!green,inner sep=2.8mm] {9.99e-01}; & \node[fill=yellow!30!white,inner sep=2.8mm,inner sep=2.8mm] {0.00e+00}; & \node[fill=yellow!30!white,inner sep=2.8mm] {-2.22e-17}; \\
            \node[fill=blue!30!white,inner sep=2.8mm] {0.00e+00}; & \node[fill=yellow!30!white,inner sep=2.8mm] {0.00e+00}; & \node[fill=white!90!green,inner sep=2.8mm,inner sep=2.8mm] {9.98e-01}; & \node[fill=yellow!30!white,inner sep=2.8mm] {0.00e+00}; \\
            \node[fill=blue!30!white,inner sep=2.8mm] {1.25e-07}; & \node[fill=yellow!30!white,inner sep=2.8mm] {-2.23e-15}; & \node[fill=yellow!30!white,inner sep=2.8mm,inner sep=2.8mm] {0.00e+00}; & \node[fill=white!90!green,inner sep=2.8mm] {9.99e-01}; \\
            };
            \end{tikzpicture}
        \subcaption{Logical PTM for 1+1 rounds of syndrome extraction without crosstalk}
    \end{subtable}
    \newline
    \vspace*{0.1cm}    
    \newline
    \begin{subtable}{.7\linewidth}\centering
        \begin{tikzpicture}[every node/.style={anchor=base,node distance=1cm, text depth=0.5ex,text height=1.5em,text width=4em}]
            \matrix [inner sep=2mm]
            {
            \node[shape=rectangle, fill={rgb:black,1;white,2},inner sep=2.8mm] {1.00e+00}; & \node[fill={rgb:black,1;white,2},inner sep=2.8mm] {-1.22e-17}; & \node[fill={rgb:black,1;white,2},inner sep=2.8mm] {-1.57e-21}; & \node[fill={rgb:black,1;white,2},inner sep=2.8mm] {8.56e-17}; \\
            \node[fill=blue!30!white,inner sep=2.8mm] {9.90e-08}; & \node[fill=white!90!green,inner sep=2.8mm] {9.99e-01}; & \node[fill=yellow!30!white,inner sep=2.8mm,inner sep=2.8mm] {-3.24e-06}; & \node[fill=yellow!30!white,inner sep=2.8mm] {-2.19e-17}; \\
            \node[fill=blue!30!white,inner sep=2.8mm] {-5.52e-11}; & \node[fill=yellow!30!white,inner sep=2.8mm] {3.17e-06}; & \node[fill=white!90!green,inner sep=2.8mm,inner sep=2.8mm] {9.98e-01}; & \node[fill=yellow!30!white,inner sep=2.8mm] {1.52e-21}; \\
            \node[fill=blue!30!white,inner sep=2.8mm] {9.50e-08}; & \node[fill=yellow!30!white,inner sep=2.8mm] {-1.24e-11}; & \node[fill=yellow!30!white,inner sep=2.8mm,inner sep=2.8mm] {-6.29e-21}; & \node[fill=white!90!green,inner sep=2.8mm] {9.99e-01}; \\
            };
            \end{tikzpicture}
        \subcaption{Logical PTM for 1+1 rounds of syndrome extraction with crosstalk strength $k=0.03$}
    \end{subtable}
    \newline
    \vspace*{0.1cm}    
    \newline
    \begin{subtable}{.7\linewidth}\centering
        \begin{tikzpicture}[every node/.style={anchor=base,node distance=1cm, text depth=0.5ex,text height=1.5em,text width=4em}]
            \matrix [inner sep=2mm]
            {
            \node[shape=rectangle, fill={rgb:black,1;white,2},inner sep=2.8mm] {0.00e+00}; & \node[fill={rgb:black,1;white,2},inner sep=2.8mm] {-6.00e-20}; & \node[fill={rgb:black,1;white,2},inner sep=2.8mm] {1.57e-21}; & \node[fill={rgb:black,1;white,2},inner sep=2.8mm] {5.11e-19}; \\
            \node[fill=blue!30!white,inner sep=2.8mm] {2.98e-08}; & \node[fill=white!90!green,inner sep=2.8mm] {9.25e-05}; & \node[fill=yellow!30!white,inner sep=2.8mm,inner sep=2.8mm] {3.24e-06}; & \node[fill=yellow!30!white,inner sep=2.8mm] {-3.19e-19}; \\
            \node[fill=blue!30!white,inner sep=2.8mm] {5.52e-11}; & \node[fill=yellow!30!white,inner sep=2.8mm] {-3.17e-06}; & \node[fill=white!90!green,inner sep=2.8mm,inner sep=2.8mm] {9.25e-05}; & \node[fill=yellow!30!white,inner sep=2.8mm] {-1.52e-21}; \\
            \node[fill=blue!30!white,inner sep=2.8mm] {2.99e-08}; & \node[fill=yellow!30!white,inner sep=2.8mm] {1.24e-11}; & \node[fill=yellow!30!white,inner sep=2.8mm,inner sep=2.8mm] {6.29e-21}; & \node[fill=white!90!green,inner sep=2.8mm] {3.44e-06}; \\
            };
            \end{tikzpicture}
        \subcaption{The difference between the above two Logical PTMs for 1+1 rounds of syndrome extraction}
    \end{subtable}
    \newline
\end{table}

\begin{table}[ht!]
    \centering
    \begin{subtable}{.7\linewidth}\centering
        \begin{tikzpicture}[every node/.style={anchor=base,node distance=1cm, text depth=0.5ex,text height=1.5em,text width=4em}]
            \matrix [inner sep=2mm]
            {
            \node[shape=rectangle, fill={rgb:black,1;white,2},inner sep=2.8mm] {1.00e+00}; & \node[fill={rgb:black,1;white,2},inner sep=2.8mm] {-4.07e-18}; & \node[fill={rgb:black,1;white,2},inner sep=2.8mm] {0.00e+00}; & \node[fill={rgb:black,1;white,2},inner sep=2.8mm] {-9.42e-18}; \\
            \node[fill=blue!30!white,inner sep=2.8mm] {2.88e-07}; & \node[fill=white!90!green,inner sep=2.8mm] {9.92e-01}; & \node[fill=yellow!30!white,inner sep=2.8mm,inner sep=2.8mm] {0.00e+00}; & \node[fill=yellow!30!white,inner sep=2.8mm] {-2.29e-13}; \\
            \node[fill=blue!30!white,inner sep=2.8mm] {0.00e+00}; & \node[fill=yellow!30!white,inner sep=2.8mm] {0.00e+00}; & \node[fill=white!90!green,inner sep=2.8mm,inner sep=2.8mm] {9.85e-01}; & \node[fill=yellow!30!white,inner sep=2.8mm] {0.00e+00}; \\
            \node[fill=blue!30!white,inner sep=2.8mm] {6.14e-06}; & \node[fill=yellow!30!white,inner sep=2.8mm] {-2.31e-13}; & \node[fill=yellow!30!white,inner sep=2.8mm,inner sep=2.8mm] {0.00e+00}; & \node[fill=white!90!green,inner sep=2.8mm] {9.92e-01}; \\
            };
            \end{tikzpicture}
        \setcounter{subtable}{3}
        \subcaption{Logical PTM for 2+1 rounds of syndrome extraction without crosstalk}
    \end{subtable}
    \newline
    \vspace*{0.1cm}    
    \newline
    \begin{subtable}{.7\linewidth}\centering
        \begin{tikzpicture}[every node/.style={anchor=base,node distance=1cm, text depth=0.5ex,text height=1.5em,text width=4em}]
            \matrix [inner sep=2mm]
            {
            \node[shape=rectangle, fill={rgb:black,1;white,2},inner sep=2.8mm] {1.00e+00}; & \node[fill={rgb:black,1;white,2},inner sep=2.8mm] {-2.45e-18}; & \node[fill={rgb:black,1;white,2},inner sep=2.8mm] {-5.37e-20}; & \node[fill={rgb:black,1;white,2},inner sep=2.8mm] {5.96e-18}; \\
            \node[fill=blue!30!white,inner sep=2.8mm] {3.47e-07}; & \node[fill=white!90!green,inner sep=2.8mm] {9.88e-01}; & \node[fill=yellow!30!white,inner sep=2.8mm,inner sep=2.8mm] {2.65e-03}; & \node[fill=yellow!30!white,inner sep=2.8mm] {2.99e-08}; \\
            \node[fill=blue!30!white,inner sep=2.8mm] {4.12e-09}; & \node[fill=yellow!30!white,inner sep=2.8mm] {-2.63e-03}; & \node[fill=white!90!green,inner sep=2.8mm,inner sep=2.8mm] {9.80e-01}; & \node[fill=yellow!30!white,inner sep=2.8mm] {-7.43e-06}; \\
            \node[fill=blue!30!white,inner sep=2.8mm] {8.89e-06}; & \node[fill=yellow!30!white,inner sep=2.8mm] {-6.75e-10}; & \node[fill=yellow!30!white,inner sep=2.8mm,inner sep=2.8mm] {7.95e-06}; & \node[fill=white!90!green,inner sep=2.8mm] {9.91e-01}; \\
            };
            \end{tikzpicture}
        \subcaption{Logical PTM for 2+1 rounds of syndrome extraction with crosstalk strength $k=0.03$}
    \end{subtable}
    \newline
    \vspace*{0.1cm}    
    \newline
    \begin{subtable}{.7\linewidth}\centering
        \begin{tikzpicture}[every node/.style={anchor=base,node distance=1cm, text depth=0.5ex,text height=1.5em,text width=4em}]
            \matrix [inner sep=2mm]
            {
            \node[shape=rectangle, fill={rgb:black,1;white,2},inner sep=2.8mm] {1.00e+00}; & \node[fill={rgb:black,1;white,2},inner sep=2.8mm] {-3.19e-18}; & \node[fill={rgb:black,1;white,2},inner sep=2.8mm] {-1.96e-19}; & \node[fill={rgb:black,1;white,2},inner sep=2.8mm] {-1.42e-18}; \\
            \node[fill=blue!30!white,inner sep=2.8mm] {3.71e-07}; & \node[fill=white!90!green,inner sep=2.8mm] {9.87e-01}; & \node[fill=yellow!30!white,inner sep=2.8mm,inner sep=2.8mm] {3.45e-03}; & \node[fill=yellow!30!white,inner sep=2.8mm] {5.09e-08}; \\
            \node[fill=blue!30!white,inner sep=2.8mm] {6.26e-09}; & \node[fill=yellow!30!white,inner sep=2.8mm] {-3.43e-03}; & \node[fill=white!90!green,inner sep=2.8mm,inner sep=2.8mm] {9.79e-01}; & \node[fill=yellow!30!white,inner sep=2.8mm] {-9.67e-06}; \\
            \node[fill=blue!30!white,inner sep=2.8mm] {9.86e-06}; & \node[fill=yellow!30!white,inner sep=2.8mm] {-9.93e-10}; & \node[fill=yellow!30!white,inner sep=2.8mm,inner sep=2.8mm] {1.04e-05}; & \node[fill=white!90!green,inner sep=2.8mm] {9.91e-01}; \\
            };
            \end{tikzpicture}
        \subcaption{Logical PTM for 2+1 rounds of syndrome extraction with crosstalk strength $k=\frac{0.23}{7}$}
    \end{subtable}
    \newline
\end{table}
\begin{table}[ht!]
    \centering
    \begin{subtable}{.7\linewidth}\centering
        \begin{tikzpicture}[every node/.style={anchor=base,node distance=1cm, text depth=0.5ex,text height=1.5em,text width=4em}]
            \matrix [inner sep=2mm]
            {
            \node[shape=rectangle, fill={rgb:black,1;white,2},inner sep=2.8mm] {1.00e+00}; & \node[fill={rgb:black,1;white,2},inner sep=2.8mm] {-2.63e-18}; & \node[fill={rgb:black,1;white,2},inner sep=2.8mm] {-2.58e-20}; & \node[fill={rgb:black,1;white,2},inner sep=2.8mm] {1.29-17}; \\
            \node[fill=blue!30!white,inner sep=2.8mm] {3.88e-07}; & \node[fill=white!90!green,inner sep=2.8mm] {9.86e-01}; & \node[fill=yellow!30!white,inner sep=2.8mm,inner sep=2.8mm] {4.39e-03}; & \node[fill=yellow!30!white,inner sep=2.8mm] {7.27e-08}; \\
            \node[fill=blue!30!white,inner sep=2.8mm] {1.02e-08}; & \node[fill=yellow!30!white,inner sep=2.8mm] {-4.36e-03}; & \node[fill=white!90!green,inner sep=2.8mm,inner sep=2.8mm] {9.78e-01}; & \node[fill=yellow!30!white,inner sep=2.8mm] {-1.23e-05}; \\
            \node[fill=blue!30!white,inner sep=2.8mm] {1.08e-05}; & \node[fill=yellow!30!white,inner sep=2.8mm] {-1.48e-09}; & \node[fill=yellow!30!white,inner sep=2.8mm,inner sep=2.8mm] {1.32e-05}; & \node[fill=white!90!green,inner sep=2.8mm] {9.91e-01}; \\
            };
            \end{tikzpicture}
        \setcounter{subtable}{6}
        \subcaption{Logical PTM for 2+1 rounds of syndrome extraction with crosstalk strength $k=\frac{0.25}{7}$}
    \end{subtable}
    \newline
    \vspace*{0.1cm}    
    \newline
    \begin{subtable}{.7\linewidth}\centering
        \begin{tikzpicture}[every node/.style={anchor=base,node distance=1cm, text depth=0.5ex,text height=1.5em,text width=4em}]
            \matrix [inner sep=2mm]
            {
            \node[shape=rectangle, fill={rgb:black,1;white,2},inner sep=2.8mm] {1.00e+00}; & \node[fill={rgb:black,1;white,2},inner sep=2.8mm] {-2.10e-18}; & \node[fill={rgb:black,1;white,2},inner sep=2.8mm] {-4.78e-20}; & \node[fill={rgb:black,1;white,2},inner sep=2.8mm] {-1.94e-18}; \\
            \node[fill=blue!30!white,inner sep=2.8mm] {3.92e-07}; & \node[fill=white!90!green,inner sep=2.8mm] {9.85e-01}; & \node[fill=yellow!30!white,inner sep=2.8mm,inner sep=2.8mm] {5.46e-03}; & \node[fill=yellow!30!white,inner sep=2.8mm] {1.09e-07}; \\
            \node[fill=blue!30!white,inner sep=2.8mm] {1.12e-08}; & \node[fill=yellow!30!white,inner sep=2.8mm] {-5.42e-03}; & \node[fill=white!90!green,inner sep=2.8mm,inner sep=2.8mm] {9.76e-01}; & \node[fill=yellow!30!white,inner sep=2.8mm] {-1.55e-05}; \\
            \node[fill=blue!30!white,inner sep=2.8mm] {1.21e-05}; & \node[fill=yellow!30!white,inner sep=2.8mm] {-2.15e-09}; & \node[fill=yellow!30!white,inner sep=2.8mm,inner sep=2.8mm] {1.65e-05}; & \node[fill=white!90!green,inner sep=2.8mm] {9.91e-01}; \\
            };
            \end{tikzpicture}
        \subcaption{Logical PTM for 2+1 rounds of syndrome extraction with crosstalk strength $k=\frac{0.27}{7}$}
    \end{subtable}
    \newline
    \vspace*{0.1cm}    
    \newline
    \begin{subtable}{.7\linewidth}\centering
        \begin{tikzpicture}[every node/.style={anchor=base,node distance=1cm, text depth=0.5ex,text height=1.5em,text width=4em}]
            \matrix [inner sep=2mm]
            {
            \node[shape=rectangle, fill={rgb:black,1;white,2},inner sep=2.8mm] {1.00e+00}; & \node[fill={rgb:black,1;white,2},inner sep=2.8mm] {-3.27e-18}; & \node[fill={rgb:black,1;white,2},inner sep=2.8mm] {2.28e-20}; & \node[fill={rgb:black,1;white,2},inner sep=2.8mm] {4.92e-18}; \\
            \node[fill=blue!30!white,inner sep=2.8mm] {4.00e-07}; & \node[fill=white!90!green,inner sep=2.8mm] {9.83e-01}; & \node[fill=yellow!30!white,inner sep=2.8mm,inner sep=2.8mm] {6.69e-03}; & \node[fill=yellow!30!white,inner sep=2.8mm] {1.49e-07}; \\
            \node[fill=blue!30!white,inner sep=2.8mm] {1.30e-08}; & \node[fill=yellow!30!white,inner sep=2.8mm] {-6.64e-03}; & \node[fill=white!90!green,inner sep=2.8mm,inner sep=2.8mm] {9.75e-01}; & \node[fill=yellow!30!white,inner sep=2.8mm] {-1.89e-05}; \\
            \node[fill=blue!30!white,inner sep=2.8mm] {1.34e-05}; & \node[fill=yellow!30!white,inner sep=2.8mm] {-2.90e-09}; & \node[fill=yellow!30!white,inner sep=2.8mm,inner sep=2.8mm] {2.03e-05}; & \node[fill=white!90!green,inner sep=2.8mm] {9.91e-01}; \\
            };
            \end{tikzpicture}
        \subcaption{Logical PTM for 2+1 rounds of syndrome extraction with crosstalk strength $k=\frac{0.29}{7}$}
    \end{subtable}
    \newline
\end{table}

\begin{table}[ht!]
    \centering
    \begin{subtable}{.7\linewidth}\centering
        \begin{tikzpicture}[every node/.style={anchor=base,node distance=1cm, text depth=0.5ex,text height=1.5em,text width=4em}]
            \matrix [inner sep=2mm]
            {
            \node[shape=rectangle, fill={rgb:black,1;white,2},inner sep=2.8mm] {1.00e+00}; & \node[fill={rgb:black,1;white,2},inner sep=2.8mm] {-1.34e-18}; & \node[fill={rgb:black,1;white,2},inner sep=2.8mm] {-5.03e-20}; & \node[fill={rgb:black,1;white,2},inner sep=2.8mm] {3.55e-19}; \\
            \node[fill=blue!30!white,inner sep=2.8mm] {3.94e-07}; & \node[fill=white!90!green,inner sep=2.8mm] {9.81-01}; & \node[fill=yellow!30!white,inner sep=2.8mm,inner sep=2.8mm] {8.07e-03}; & \node[fill=yellow!30!white,inner sep=2.8mm] {1.93e-07}; \\
            \node[fill=blue!30!white,inner sep=2.8mm] {1.48e-08}; & \node[fill=yellow!30!white,inner sep=2.8mm] {-8.01e-03}; & \node[fill=white!90!green,inner sep=2.8mm,inner sep=2.8mm] {9.73e-01}; & \node[fill=yellow!30!white,inner sep=2.8mm] {-2.27e-05}; \\
            \node[fill=blue!30!white,inner sep=2.8mm] {1.47e-05}; & \node[fill=yellow!30!white,inner sep=2.8mm] {-4.10e-09}; & \node[fill=yellow!30!white,inner sep=2.8mm,inner sep=2.8mm] {2.45e-05}; & \node[fill=white!90!green,inner sep=2.8mm] {9.91e-01}; \\
            };
            \end{tikzpicture}
        \setcounter{subtable}{9}
        \subcaption{Logical PTM for 2+1 rounds of syndrome extraction with crosstalk strength $k=\frac{0.31}{7}$}
    \end{subtable}
    \newline
    \vspace*{0.1cm}    
    \newline
    \begin{subtable}{.7\linewidth}\centering
        \begin{tikzpicture}[every node/.style={anchor=base,node distance=1cm, text depth=0.5ex,text height=1.5em,text width=4em}]
            \matrix [inner sep=2mm]
            {
            \node[shape=rectangle, fill={rgb:black,1;white,2},inner sep=2.8mm] {1.00e+00}; & \node[fill={rgb:black,1;white,2},inner sep=2.8mm] {-2.20e-18}; & \node[fill={rgb:black,1;white,2},inner sep=2.8mm] {3.40e-20}; & \node[fill={rgb:black,1;white,2},inner sep=2.8mm] {-4.13e-18}; \\
            \node[fill=blue!30!white,inner sep=2.8mm] {4.22e-07}; & \node[fill=white!90!green,inner sep=2.8mm] {9.80e-01}; & \node[fill=yellow!30!white,inner sep=2.8mm,inner sep=2.8mm] {9.61e-03}; & \node[fill=yellow!30!white,inner sep=2.8mm] {2.91e-07}; \\
            \node[fill=blue!30!white,inner sep=2.8mm] {2.06e-08}; & \node[fill=yellow!30!white,inner sep=2.8mm] {-9.56e-03}; & \node[fill=white!90!green,inner sep=2.8mm,inner sep=2.8mm] {9.71e-01}; & \node[fill=yellow!30!white,inner sep=2.8mm] {-2.70e-05}; \\
            \node[fill=blue!30!white,inner sep=2.8mm] {1.42e-05}; & \node[fill=yellow!30!white,inner sep=2.8mm] {-5.26e-09}; & \node[fill=yellow!30!white,inner sep=2.8mm,inner sep=2.8mm] {2.92e-05}; & \node[fill=white!90!green,inner sep=2.8mm] {9.91e-01}; \\
            };
            \end{tikzpicture}
        \subcaption{Logical PTM for 2+1 rounds of syndrome extraction with crosstalk strength $k=\frac{0.33}{7}$}
    \end{subtable}
    \newline
    \vspace*{0.1cm}    
    \newline
    \begin{subtable}{.7\linewidth}\centering
        \begin{tikzpicture}[every node/.style={anchor=base,node distance=1cm, text depth=0.5ex,text height=1.5em,text width=4em}]
            \matrix [inner sep=2mm]
            {
            \node[shape=rectangle, fill={rgb:black,1;white,2},inner sep=2.8mm] {1.00e+00}; & \node[fill={rgb:black,1;white,2},inner sep=2.8mm] {-1.76e-18}; & \node[fill={rgb:black,1;white,2},inner sep=2.8mm] {1.61e-19}; & \node[fill={rgb:black,1;white,2},inner sep=2.8mm] {-3.42e-19}; \\
            \node[fill=blue!30!white,inner sep=2.8mm] {4.28e-07}; & \node[fill=white!90!green,inner sep=2.8mm] {9.78e-01}; & \node[fill=yellow!30!white,inner sep=2.8mm,inner sep=2.8mm] {1.13e-02}; & \node[fill=yellow!30!white,inner sep=2.8mm] {3.23e-07}; \\
            \node[fill=blue!30!white,inner sep=2.8mm] {2.29e-08}; & \node[fill=yellow!30!white,inner sep=2.8mm] {-1.13e-02}; & \node[fill=white!90!green,inner sep=2.8mm,inner sep=2.8mm] {9.69e-01}; & \node[fill=yellow!30!white,inner sep=2.8mm] {-3.21e-05}; \\
            \node[fill=blue!30!white,inner sep=2.8mm] {1.39e-05}; & \node[fill=yellow!30!white,inner sep=2.8mm] {-6.70e-09}; & \node[fill=yellow!30!white,inner sep=2.8mm,inner sep=2.8mm] {3.44e-05}; & \node[fill=white!90!green,inner sep=2.8mm] {9.91e-01}; \\
            };
            \end{tikzpicture}
        \subcaption{Logical PTM for 2+1 rounds of syndrome extraction with crosstalk strength $k=0.05$}
    \end{subtable}
    \newline
\end{table}
\end{section}

\end{document}